\documentclass[aps, prl, superscriptaddress, preprintnumbers, floatfix, longbibliography,twocolumn,10pt]{revtex4-2}

\usepackage{graphicx,amsfonts,amsmath,amssymb,amstext}
\usepackage{float,wrapfig}
\usepackage{subfigure, psfrag}
\usepackage{dsfont,bm}
\usepackage{color}
\usepackage[colorlinks=true,urlcolor=blue,citecolor=blue,linkcolor=blue]{hyperref}

\definecolor{purple}{rgb}{0.8,0,0.6}
\definecolor{darkgreen}{rgb}{0.00,0.6,0.00}

\extrafloats{100}

\begin{document}

\title{Anomalous Electromagnetic Field Penetration in a Weyl or Dirac Semimetal}
\date{April 6, 2022}

\author{P.~O.~Sukhachov}
\email{pavlo.sukhachov@yale.edu}
\affiliation{Department of Physics, Yale University, New Haven, Connecticut 06520, USA}

\author{L.~I.~Glazman}
\affiliation{Department of Physics, Yale University, New Haven, Connecticut 06520, USA}

\begin{abstract}
The current response to an electromagnetic field in a Weyl or Dirac semimetal becomes nonlocal due to the chiral anomaly activated by an applied static magnetic field. The nonlocality develops under the conditions of the normal skin effect and is related to the valley charge imbalance generated by the joint effect of the electric field of the impinging wave and the static magnetic field. We elucidate the signatures of this nonlocality in the transmission of electromagnetic waves. The signatures include enhancement of the transmission amplitude and its specific dependence on the wave's frequency and the static magnetic field strength.
\end{abstract}

\maketitle

{\it Introduction.---}
A salient feature of Weyl and Dirac materials is the possibility to realize the chiral anomaly due to their relativisticlike electronic spectra in the vicinity of the band-touching nodal points. As was pointed out in Ref.~\cite{Nielsen:1983}, this is an analog of the Adler-Bell-Jackiw axial anomaly in relativistic physics~\cite{Adler:1969,BJ:1969}. The chiral Adler-Bell-Jackiw anomaly was first observed in Weyl superfluid $^3$He-A~\cite{Bevan-Volovik:1997}. In the solid-state physics setting, the anomaly may lead to a negative magnetoresistance in the direction parallel to the applied magnetic field. Interest in the manifestations of the chiral anomaly in the electron transport flared up after the discovery of Weyl semimetals~\cite{Hasan-Huang:rev-2017,Armitage:rev-2018,GMSS:book}. The kinetic theory of negative magnetoresistance in direct current (dc) transport was fleshed out~\cite{Son-Spivak:2013} and its dependence on the electron spectra and relaxation times was elucidated. A negative magnetoresistance was indeed observed in Dirac ({\sl e.g.,} Na$_3$Bi, Cd$_3$As$_2$, and ZrTe$_5$) and Weyl ({\sl e.g.,} transition metal monopnictides TaAs, NbAs, TaP, and NbP) semimetals (see Refs.~\cite{Hosur-Qi:rev-2013,Burkov:rev-2015,Gorbar:2017lnp,Hu-Mao:rev-2019,Ong-Liang:rev-2020} for reviews on anomalous transport properties). However, it was soon realized that the observation of the negative magnetoresistivity alone is not sufficient to claim the realization of the chiral anomaly. Among the effects that can mimic the anomaly are current jetting~\cite{Reis-Hassinger:2016,Liang-Ong:2018} due to an inhomogeneous distribution of the electric current in materials with high mobility and electron scattering on long-range ionic impurities~\cite{Goswami-DasSarma:2015}.

It was suggested in Ref.~\cite{Burkov:2018-OptCond} to use frequency as an additional control ``knob'' to investigate the effects of the chiral anomaly while circumventing the current jetting: in the presence of a magnetic field, the anomaly results in a Drude-like contribution to the conductivity. The width of the corresponding low-frequency peak in the linear alternating current (ac) response to a spatially uniform electric field is determined by the internode relaxation rate. The latter rate is usually small compared with the intranode relaxation rate, so the anomalous conductivity peak is fairly narrow. The tendency toward peak narrowing was seen in the contact-less measurements of the transmission amplitude of an electromagnetic field through a Cd$_3$As$_2$ film~\cite{Cheng-Armitage:2019}.

The electric field of the wave penetrating a material, however, is nonuniform due to the skin effect. This raises a question regarding the influence of chiral anomaly on the transmission of an electromagnetic wave across a film made of a Weyl or Dirac conductor.

We demonstrate in this Letter that an application of a magnetic field parallel to the surface of a Weyl or Dirac conductor activate the chiral anomaly and may result in a nonlocal current response to an impinging electromagnetic wave. We emphasize that this nonlocal response develops under the conditions corresponding to the normal skin effect. The latter is thought to be adequate for materials with the electron mean free path shorter than the electromagnetic field penetration depth~\cite{Abrikosov:book-1988}. A new element brought by the topological electronic band structure is the valley charge imbalance. It is activated via the chiral anomaly by the joint effect of the electric field of the impinging wave, active within the skin layer, and a static magnetic field. The valley charge imbalance preserves the local charge neutrality and therefore is not suppressed by screening. This property allows the imbalance to diffuse beyond the skin depth, deeper into the sample. The accompanying chiral magnetic effect~\cite{Vilenkin:1980,Fukushima:2008} current represents the nonlocal response to the electric field of the impinging wave and facilitates its anomalous penetration similar to a dc nonlocal transport~\cite{Parameswaran-Vishwanath:2014,Zhang-Xiu:2017,Boer-Brinkman:2019}.

\begin{figure}[t]
\centering
\includegraphics[width=0.45\textwidth]{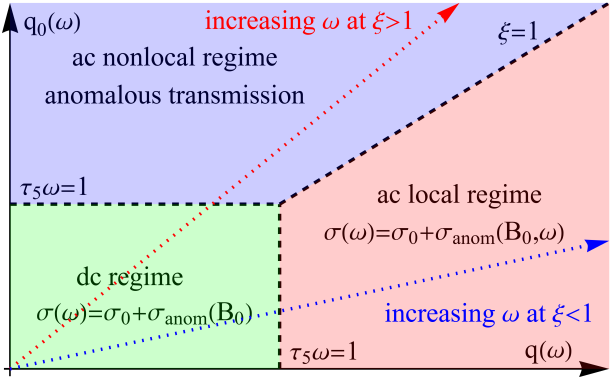}
\caption{
The schematic representation of the current response regimes discussed in this work. Here $q_0(\omega) =1/\delta(\omega)= \sqrt{2\pi \sigma_0 \omega}/c$ is inverse of the skin depth, $\sigma_0$ is the static Drude conductivity, $\omega$ is the angular frequency of the impinging wave, $q(\omega) = \sqrt{\omega/(2D)}$ is inverse of the diffusion length, $D$ is the diffusion coefficient, $\xi = q_0(\omega)/q(\omega)= \sqrt{4\pi \sigma_0 D}/c$ is the frequency-independent parameter quantifying the nonlocality of the response, and $1/\tau_5$ is the effective internode scattering rate. In addition, we assume a short intranode scattering time $\tau$, {\sl i.e.}, $\omega \tau\ll1$. The transmission of electromagnetic waves is described via the standard expressions for the normal skin effect~\cite{Abrikosov:book-1988} with a conductivity modified by the chiral anomaly, $\sigma(\omega) = \sigma_0 +\sigma_{\rm anom}(B_0,\omega)$, in the dc and ac local ($\xi<1$) regimes; see Eq.~(\ref{Eslab2-local}) for the transmitted electric field. In the ac nonlocal regime ($\xi>1$), it is possible to achieve an enhancement of the electromagnetic wave penetration depth; see Eqs.~(\ref{Eslab2-short}) and (\ref{Eslab2-long}).
}
\label{fig:diagramm}
\end{figure}

The three main regimes of the current response including dc, ac local, and ac nonlocal regimes are schematically illustrated in Fig.~\ref{fig:diagramm}. In this work, unlike the existing studies ({\sl e.g.}, Ref.~\cite{Cheng-Armitage:2019}) of the chiral anomaly performed in the local regimes (see the blue dotted line in Fig.~\ref{fig:diagramm}) we focus on the \emph{ac nonlocal} regime with a spatial dispersion of the conductivity (see the red dotted line in Fig.~\ref{fig:diagramm}).

{\it Model and key equations.---} To study the transmission of electromagnetic waves, we consider a film of a Dirac or time-reversal symmetric Weyl semimetal~\footnote{The main qualitative results of our study should hold for a time-reversal symmetry broken Weyl semimetal as well. However, such systems have other phenomena that also affect the propagation and reflection of electromagnetic waves, {\sl e.g.}, anomalous Kerr and Faraday effects~\cite{Kargarian-Trivedi:2015}. These phenomena are ignored in this work where we focus on the effects of the chiral anomaly.} with the thickness $L$ along the $z$ direction. We assume the normal incidence of the incoming ($z\leq 0$) wave with an electric field $\mathbf{E}_{\rm in}(t,z)=\mathbf{E}_{\rm in} e^{i(kz-\omega t)}$, where $\omega$ is the angular frequency and $k=\omega/c$ is the wave vector. A portion of the incoming field $\mathbf{E}_{\rm r}(t,z)$ is reflected from the surface and a portion $\mathbf{E}_{\rm out}(t,z)$ is transmitted across the film. The in-medium field $\mathbf{E}(t,z)$ satisfies the standard system of Maxwell's equations. To close it, one needs to evaluate the current density as a response to the electric field. This (generally nonlocal) linear response is controlled by the electron kinetics. In building the kinetic theory of a Weyl or Dirac semimetal, we assume that the characteristic intranode relaxation times are much shorter than the internode ones in accordance with experiments, see, {\sl e.g.}, Refs.~\cite{Cheng-Armitage:2019,Jadidi-Drew-TaAs:2019}. In addition, the intranode scattering rates are assumed to be much larger than the frequency of the electromagnetic field. To activate the chiral anomaly, we include a static uniform magnetic field $\mathbf{B}_0$, which is applied parallel to the surface and is classically weak~\footnote{The cyclotron frequency for the classically-weak magnetic field is smaller than the intranode scattering rate, {\sl i.e.}, $\omega_c\ll \mbox{min}\left\{1/\tau_{\alpha,\alpha}\right\}$, where $1/\tau_{\alpha,\alpha}$ is the intranode scattering rate for Weyl node $\alpha$.}. Under this condition, $\mathbf{B}_0$ does not affect the diffusive electron dynamics while introducing an anomalous term into the partial current density $\mathbf{j}_{\alpha}(t,z)$ produced by electrons of node $\alpha$~\footnote{We disregard the contribution of the Fermi arc surface states in the current. As is estimated in Ref.~\cite{Chen-Belyanin:2019}, the relative contribution of the Fermi arcs to the surface impedance is negligible if the frequency of the impinging wave is much smaller than the plasmon resonance frequency, which is indeed the case in our study.},
\begin{equation}
\label{CKT-equation-node-j}
\mathbf{j}_{\alpha}(t,z) = \sigma_{\alpha}\mathbf{E}(t,z) -D_{\alpha}\bm{\nabla}N_{\alpha}(t,z)-\mathbf{v}_{\Omega,\alpha}N_{\alpha}(t,z).
\end{equation}
Here $N_{\alpha}(t,z)$ is the perturbed partial (or valley) electron charge density at node $\alpha$ and $\mathbf{v}_{\Omega,\alpha}$ is the anomalous velocity associated with the flux $\chi_{\alpha}$ of the Berry curvature; $D_{\alpha}$ and $\sigma_{\alpha}=e^2 \nu_{\alpha} D_{\alpha}$ are the respective diffusion constant and partial electric conductivity. In terms of $\chi_\alpha$ and the Fermi level density of states $\nu_\alpha$ of electrons around node $\alpha$, the anomalous velocity is $\mathbf{v}_{\Omega,\alpha}= \chi_{\alpha}e\mathbf{B}_0/\left(4\pi^2 \hbar^2 c \nu_{\alpha}\right)$. While the first two terms in Eq.~(\ref{CKT-equation-node-j}) correspond to the conventional intranode diffusion current, the last term describes the chiral magnetic effect current~\cite{Vilenkin:1980,Fukushima:2008} after summing over all nodes.

The kinetic equation in the diffusive approximation is
\begin{eqnarray}
\label{CKT-equation-n-chi}
&&\partial_t N_{\alpha}(t,z) +\bm{\nabla}\cdot\mathbf{j}_{\alpha}(t,z) = -\sum_{\beta}^{N_{W}}T_{\alpha,\beta}N_{\beta}(t,z) \nonumber\\
&&-e^2\nu_{\alpha} \mathbf{v}_{\Omega,\alpha}\cdot\mathbf{E}(t,z);
\end{eqnarray}
see the Supplemental Material~\cite{SM} and, {\sl e.g.}, Refs.~\cite{Burkov:2014b,Parameswaran-Vishwanath:2014,Burkov:rev-2015} for details. The terms on the left-hand side of Eq.~(\ref{CKT-equation-n-chi}) correspond to the conventional continuity equation in each of the nodes. On the right-hand side, the shorthand notation $T_{\alpha,\beta} = \delta_{\alpha,\beta} \sum_{\gamma}^{N_{W}}1/\tau_{\alpha,\gamma} -1/\tau_{\beta,\alpha}$ in the term responsible for the internode scattering in the relaxation time approximation was introduced. Here $N_{W}$ is a number of Weyl nodes and $1/\tau_{\alpha,\beta}$ are the scattering rates between nodes $\alpha$ and $\beta$. Finally, the last term in Eq.~(\ref{CKT-equation-n-chi}) corresponds to the chiral anomaly. It is important to note that the total electric charge $\sum_{\alpha}^{N_{W}} N_{\alpha}(t,z)$ is conserved by the collision integral and the chiral anomaly. In addition, the transverse field, $\bm{\nabla}\cdot\mathbf E(z)=0$, in Eq.~(\ref{CKT-equation-n-chi}) does not violate the electric charge neutrality.

Since the time dependence of fields, currents, and densities is given by the same prefactor $e^{-i\omega t}$, we combine Eqs.~(\ref{CKT-equation-node-j}) and (\ref{CKT-equation-n-chi}) as
\begin{eqnarray}
\label{CKT-equation-n-chi-1}
&&\sum_{\beta}^{N_{W}}\left[\frac{T_{\alpha,\beta}}{D_{\alpha}} - 2iq_{\alpha}^2(\omega) \delta_{\alpha,\beta} -\delta_{\alpha,\beta}\partial_z^2\right] N_{\beta}(z) \nonumber\\
&&=-\frac{e^2\nu_{\alpha}}{D_{\alpha}} \mathbf{v}_{\Omega,\alpha}\cdot\mathbf{E}(z),
\end{eqnarray}
where $q_{\alpha}(\omega)=\sqrt{\omega/(2D_{\alpha})}$ is the inverse of the diffusion length. Finally, neglecting the displacement current for $\omega\ll\sigma_0$ with $\sigma_0=\sum_{\alpha}^{N_{W}}\sigma_{\alpha}$ being the static conductivity, Maxwell's equations for the transverse components of the electric field together with the equation for current [Eq.~(\ref{CKT-equation-node-j})] are brought to the following form:
\begin{equation}
\label{CKT-sol-Maxwell-general}
\left[\partial_z^2 + 2i q_0^2(\omega) \right] \mathbf{E}(z) = \frac{4\pi i \omega}{c^2}  \sum_{\alpha}^{N_{W}} \mathbf{v}_{\Omega,\alpha} N_{\alpha}(z),
\end{equation}
where $q_0(\omega) =\sqrt{2\pi \sigma_0 \omega}/c$ is the inverse of the skin depth.

In order to form a complete system for the transverse electric field $\mathbf{E}(z)$ and the valley charge densities $N_{\alpha}(z)$, Eqs.~(\ref{CKT-equation-n-chi-1}) and (\ref{CKT-sol-Maxwell-general}) should be amended with boundary conditions. We use the standard boundary conditions for electromagnetic fields, {\sl i.e.}, we require the continuity of the tangential component of the electric fields and their derivatives~\footnote{The continuity of the derivatives of the tangential components of electric fields at the surfaces follows from the continuity of the tangential components of the magnetic fields.} at $z=0,L$. As for the densities, we consider two types of phenomenological boundary conditions:
\begin{equation}
\label{CKT-sol-n-alpha-BC}
\mbox{(i)}\,\, N_{\alpha}(z=0,L)=0 \,\, \mbox{and} \,\, \mbox{(ii)}\,\, \partial_z N_{\alpha}(z=0,L)=0.
\end{equation}
These two conditions correspond, respectively, to the limits of fast and no internode relaxation at the boundary.

{\it Transmission of electromagnetic waves.---}
A finite anomalous velocity $\mathbf{v}_{\Omega,\alpha}$ emerging at $B_0\neq 0$ couples the electric field $\mathbf{E}(z)$ of the wave to the diffusion of partial densities $N_\alpha(z)$. The spectrum of the diffusion length scales can be found by solving the eigenvalue problem for the coupled set of the diffusion equations; see Eq.~(\ref{CKT-equation-n-chi-1}) for a diffusion equation at node $\alpha$. In general, the spectrum of the diffusion lengths depends on the internode relaxation rates. However, in the limit of $\omega$ being high compared with the characteristic value $1/\tau_5$ of the internode scattering rates, the diffusion equations decouple from each other, and the diffusion lengths are quantified by $1/q_\alpha(\omega)$. We note that the ratio $\xi_{\alpha}= q_0(\omega)/q_{\alpha}(\omega)= \sqrt{4\pi \sigma_0 D_{\alpha}}/c$ is defined solely by the material properties and is independent of $\omega$. The anomalous penetration of the field is driven by the largest among $\xi_\alpha$. Aiming at a strong anomalous effect, we assume $\xi_\alpha\gg 1$ for all $\alpha$ and consider films of thickness far exceeding the normal-skin penetration depth $L\gg 1/q_0(\omega)$.

It is convenient to separate the electric field into two components, $\mathbf{E}(z)=\mathbf{E}_\parallel(z)+\mathbf{E}_\perp(z)$, parallel and normal to $\mathbf{B}_0$, respectively. The anomaly affects only the former one, while $|E_\perp(z)|\propto e^{-Lq_0(\omega)}$ is independent of $B_0$. When evaluating $\mathbf{E}_\parallel(z)$, we focus on the most practical case of weak coupling between $\mathbf{E}_\parallel(z)$ and $N_\alpha(z)$. This allows us to solve Eqs. (\ref{CKT-equation-n-chi-1}) and (\ref{CKT-sol-Maxwell-general}) iteratively in $\mathbf{v}_{\Omega,\alpha}$ by starting with $E_\parallel^{(0)}(z) = (1-i)(\omega/c)\,e^{-z q_0(\omega)} e^{iz q_0(\omega)}E_{\parallel\rm in}/q_0(\omega)$
at $L-z\gg 1/q_0(\omega)$ within the film; the corresponding outgoing field follows from the boundary conditions and reads $E_{\parallel\rm out}^{(0)}(z=L)=2(1-i)(\omega/c) \, e^{-L q_0(\omega)} e^{iL q_0(\omega)}E_{\parallel\rm in}/q_0(\omega)$.
Being substituted into the right-hand side of Eq.~(\ref{CKT-equation-n-chi-1}), $E_\parallel^{(0)}(z)$ creates a source exciting valley charge density imbalance.
The resulting solution $N^{(1)}_\alpha(z)\propto v_{\Omega,\alpha}$ reads~\cite{SM} as
\begin{equation}
\label{n-2-long-i}
N_{\alpha}^{(1)}(z) = -i\frac{e^2 \nu_{\alpha} v_{\Omega,\alpha}}{2q_0^2(\omega) D_{\alpha}} \frac{\sin{\left[(1+i)(L-z)q_{\alpha}(\omega)\right]}}{\sin{\left[(1+i)q_{\alpha}(\omega)L\right]}} E_\parallel^{(0)}(0)
\end{equation}
for the Dirichlet boundary conditions [Eq.~(\ref{CKT-sol-n-alpha-BC})]. In solving Eq.~(\ref{CKT-equation-n-chi-1}), we assumed a highly nonlocal regime, $\xi_\alpha\gg 1$, and considered $z\gg 1/q_0(\omega)$.

Lastly, we use Eq.~(\ref{n-2-long-i}) on the right-hand side of Eq.~(\ref{CKT-sol-Maxwell-general}) to find the anomalous correction $E_\parallel^{(2)}(z)\propto v_{\Omega,\alpha}^2$ to the electric field. The solution to Eq.~(\ref{CKT-sol-Maxwell-general}) is simplified by a slow spatial variation of the partial densities, $1/q_\alpha(\omega)=\xi_\alpha/q_0(\omega)\gg 1/q_0(\omega)$, allowing us to write
\begin{eqnarray}
\label{Eslab2}
&&E_\parallel^{(2)}(z) = \frac{1}{\sigma_0} \sum_{\alpha}^{N_{W}} v_{\Omega,\alpha} \Bigg[N_{\alpha}^{(1)}(z)
-\frac{1+i}{2q_0(\omega)} \nonumber\\
&&\times e^{-(L-z) q_0(\omega)} e^{i(L-z) q_0(\omega)} \partial_zN_{\alpha}^{(1)}(z=L)\Bigg].
\end{eqnarray}
This form is valid for either of the two boundary conditions for $N_\alpha(z)$. The outgoing field follows from the continuity of the tangential components of the electric field, {\sl i.e.}, $E_{\parallel\rm out}^{(2)}(z=L) =E_\parallel^{(2)}(z=L)$.

We consider two characteristic cases of a thick film, $L\gg 1/q_{\alpha}(\omega)$, and a thin film, $L\ll 1/q_{\alpha}(\omega)$, compared with the diffusion lengths. In the former case, the partial charge density decays exponentially with $z$. Using Eq.~(\ref{Eslab2}), we find the following transmitted electric field:
\begin{eqnarray}
\label{Eslab2-short}
&&E_{\parallel\rm out}(t,z=L) =\!
2\sqrt{\frac{\omega}{\pi \sigma_0}} \Bigg[e^{-L/\delta(\omega)}\cos\!\left(\!\frac{L}{\delta(\omega)}-\frac{\pi}{4}-\omega t\!\right)
\nonumber\\
&&-\!\sum_\alpha^{N_{W}}\!\frac{g_\alpha}{\xi^{3}_\alpha}\frac{B^2_0}{B^2_\alpha(\omega)}e^{-L/\left[\xi_\alpha\delta(\omega)\right]} \!\cos\!\left(\!\frac{L}{\xi_\alpha\delta(\omega)}+\frac{\pi}{4}-\omega t\!\right)
\!\!\Bigg]\! E_{\parallel\rm in},
\nonumber\\
\end{eqnarray}
where $g_{\alpha}=1$ for $N_{\alpha}^{(1)}(z=0,L)=0$ and $g_{\alpha}=\xi_{\alpha}^2$ for $\partial_z N_{\alpha}^{(1)}(z=0,L)=0$, respectively. For clarity, in Eq.~(\ref{Eslab2-short}), we restored the real part for the fields, used the conventional definition for the normal-skin depth, $\delta(\omega)=c/\sqrt{2\pi \sigma_0 \omega}$, and introduced the characteristic magnetic field $B_\alpha(\omega)=4\pi\Phi_0\hbar\sqrt{\omega\nu_\alpha \sum_\beta^{N_{W}}\nu_\beta D_\beta}$, which depends on the electronic properties of the material and frequency. In writing $B_\alpha(\omega)$, we used the explicit expression for $\mathbf{v}_{\Omega,\alpha}$ and $\sigma_0=e^2\sum_\alpha^{N_W}\nu_\alpha D_\alpha$ for the Drude conductivity; $\Phi_0=\pi\hbar c/e$ is the magnetic flux quantum. While the terms in Eq.~(\ref{Eslab2-short}) representing the conventional and anomalous components of the transmitted field both decay exponentially with the film thickness, the respective penetration depths are vastly different at $\xi_\alpha\gg 1$.

In the case of a thin film, $L\ll 1/q_{\alpha}(\omega)$, the partial charge, which is created in the skin layer, spreads over the entire thickness of the film $L$ due to diffusion. Substituting the proper limit of Eq.~(\ref{n-2-long-i}) that defines $N_{\alpha}^{(1)}(z)$ into Eq.~(\ref{Eslab2}), we find
\begin{eqnarray}
\label{Eslab2-long}
&&E_{\parallel\rm out}(t,z=L) =\! 2\sqrt{\frac{\omega}{\pi \sigma_0}} \Bigg[e^{-L/\delta(\omega)}\cos\!\left(\!\frac{L}{\delta(\omega)}-\frac{\pi}{4}-\omega t\!\right)\nonumber\\
&&-\frac{1}{2\sqrt{2}}\frac{\delta(\omega)}{L} \sum_\alpha^{N_{W}} \frac{g_{\alpha}}{\xi_{\alpha}^2} \frac{B^2_0}{B^2_\alpha(\omega)} \sin{(\omega t)}\Bigg] E_{\parallel\rm in}.
\end{eqnarray}
As expected, the anomalous correction to the outgoing electric field (the second term) acquires a $\propto 1/L$ scaling with the film thickness. In the case of the Dirichlet boundary conditions ($g_{\alpha}=1$), there is an additional small prefactor $1/\xi_{\alpha}^2$ that originates from the suppression of $N_{\alpha}^{(1)}(z)$ near the boundaries. Such suppression is absent for the Neumann boundary conditions ($g_{\alpha}=\xi_{\alpha}^2$) where a uniform partial charge density is allowed~\cite{SM}.

To contrast the results for the local and nonlocal regimes, we also present the transmitted field at $\xi_{\alpha}\ll1$. It can be obtained by introducing the anomalous correction to the electric conductivity in the standard expression for the normal skin effect; see the Supplemental Material~\cite{SM} for details. In the leading order in $B_0$, we have
\begin{eqnarray}
\label{Eslab2-local}
&&E_{\rm \parallel out}(t,z=L) = 2\sqrt{\frac{\omega}{\pi \sigma_0}} e^{-L/\delta(\omega)} \Bigg[\!\cos\!\left(\!\frac{L}{\delta(\omega)}-\frac{\pi}{4}-\omega t\!\right) \nonumber\\
&&-\frac{1}{\sqrt{2}} \frac{L}{\delta(\omega)} \sum_{\alpha}^{N_{W}} \frac{B^2_0}{B^2_\alpha(\omega)} \cos\!\left(\!\frac{L}{\delta(\omega)}-\omega t\!\right)
\Bigg] E_{\parallel\rm in},
\end{eqnarray}
where, as in the case of the nonlocal response, we neglected the internode scattering. As one can see by comparing Eqs.~(\ref{Eslab2-short}), (\ref{Eslab2-long}), and (\ref{Eslab2-local}), the scaling of the anomalous parts of the transmitted fields with frequency is qualitatively different and might be used to distinguish nonlocal and local response regimes even if material parameters are not known {\sl a priori}. Furthermore, it is straightforward to check~\cite{SM} that the amplitude of the transmitted field in the local regime always decreases with the magnetic field. On the other hand, interference between the anomalous and the regular terms in Eq.~(\ref{Eslab2-short}) or (\ref{Eslab2-long}) may lead to an enhancement of the transmitted field at $B_0\neq0$.

{\it Estimates for a model with symmetric Weyl nodes.---}
To provide estimates of the proposed effects, we consider a simplified model with $N_{W}$ Weyl nodes forming well-separated from each other symmetric pairs. Each pair consists of nodes carrying opposite topological charges. We assume the electron dispersion around each of the nodes to be linear, with the same parameters $\nu_\alpha\to\nu$ and $D_\alpha\to D$. This allows us to introduce the node-independent electron mean free path $\ell=v_F\tau$ with the intranode relaxation time $\tau$, and replace $\xi_\alpha\to\xi$. With these simplifications, we reformulate the condition of the normal skin effect, $\ell\ll\delta(\omega)$, as $\xi\sqrt{\omega\tau}\ll 1$. Therefore, our approximations are valid for the following double constraint on $\xi$: $1\ll\xi\ll 1/\sqrt{\omega\tau}$. The lower constraint on frequency $\omega$ comes from the internode relaxation rate. In our model, the corresponding rate, $1/\tau_5$, comes from relaxation within $(\alpha,-\alpha)$ pairs. At the lower limit for frequency, $\omega\sim 1/\tau_5$, the range for $\xi$ is limited from above by $\sqrt{\tau_5/\tau}$; see also the Supplemental Material~\cite{SM}.

The magnitude of the anomalous correction to the transmitted field is controlled by the ratio $B_0/B_\alpha(\omega)$ in Eqs.~(\ref{Eslab2-short}), (\ref{Eslab2-long}), and (\ref{Eslab2-local}). In the simplified model, there is no dependence on $\alpha$, and we are able to transform $B_\alpha(\omega)\to B^\star(\omega)=(4/\sqrt{3})B_{\rm uq}\sqrt{N_{W}\omega\tau}$. Here $B_{\rm uq}$ is the magnetic field at which the ultra-quantum limit ({\sl i.e.}, only the lowest Landau level is populated) is reached. At the lowest frequencies, $\omega\sim 1/\tau_5$, the characteristic field is $B^\star\sim B_{\rm uq}\sqrt{N_{W}\tau/\tau_5}$.

To flesh out the estimates, we use some of the parameters of the Weyl semimetal TaAs~\footnote{While we use the parameters of the Weyl semimetal TaAs, other materials with a simpler band structure might be used to observe the proposed anomalous nonlocal effect. For example, we mention the Dirac semimetal Cd$_3$As$_2$~\cite{Borisenko:2014,Liu-Chen-Cd3As2:2014,Neupane-Hasan-Cd3As2:2014} and the Weyl semimetal EuCd$_2$As$_2$~\cite{Wang-Canfield:2019,Soh-Boothroyd:2019,Ma-Shi:2019}. Compared to TaAs, they have a simpler band structure with only two Dirac points and Weyl nodes, respectively.} derived from Refs.~\cite{Arnold-Felser:2016b,Zhang-Hasan-TaAs:2016}: $N_{W}=24$, the Fermi velocity $v_{F}\approx 3\times 10^7~\mbox{cm/s}$, the Fermi level (measured from a node) $\mu\approx 20~\mbox{meV}$, and the ratio $\tau_5/\tau\approx 158$. We estimate $B_{\rm uq}\approx 3.5~\mbox{T}$, the upper limit $\xi\sim 13$ for the range of $\xi$, and the lower limit $B^\star\sim 1.4~\mbox{T}$ for $B^\star\sim B_{\rm uq}\sqrt{N_{W}\tau/\tau_5}$. The above estimates depend on the ratio $\tau_5/\tau$, but not separately on any of these times. To get $\xi\gtrsim 1$, however, one needs $\tau\gtrsim 10~\mbox{ps}$; this is about $25$ times higher than the value $\tau\approx 0.38~\mbox{ps}$ reported in Refs.~\cite{Arnold-Felser:2016b,Zhang-Hasan-TaAs:2016}. One may expect the above quoted ratio $\tau_5/\tau$ to persist for cleaner samples if both $\tau$ and $\tau_5$ are limited by scattering off the same defects. Lastly, at $\tau\sim 10~\mbox{ps}$, fields $B_0\lesssim 0.02~\mbox{T}$ satisfy the condition of a classically weak field.

\begin{figure}[t]
\centering
\includegraphics[width=0.45\textwidth]{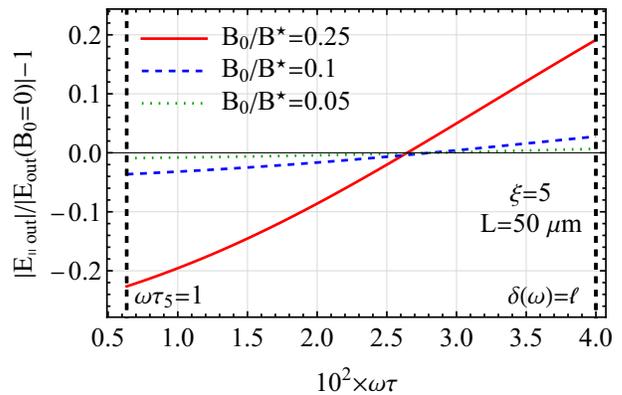}
\caption{
The dependence of the relative field amplitude $\left|E_{\rm \parallel out}\right|/\left|E_{\rm out}(B_0=0)\right|-1$ on frequency for a few values of the magnetic field. We used Eq.~(\ref{Eslab2-long}) to plot the results in the nonlocal ($\xi=5$) regime. Black vertical dashed lines are the boundaries of the parameter region where the nonlocal regime under the conditions of the normal skin effect is realized. We used Neumann boundary conditions and set $B^\star= B_{\rm uq}\sqrt{N_{W}\tau/\tau_5}$, $B_{\rm uq}=c\mu^2/(2e \hbar v_F^2)$, and $L=50~\mu\mbox{m}$; other parameters are given in the text.
}
\label{fig:anisotropy-2}
\end{figure}

We illustrate the dependence of the relative field amplitude $\left|E_{\rm \parallel out}\right|/\left|E_{\rm out}(B_0=0)\right|-1$ on frequency in Fig.~\ref{fig:anisotropy-2} for the nonlocal regime. Since cyclotron motion does not affect the conductivity along the direction of a nonquantizing magnetic field ($B_0\ll B_{\rm uq}$) for spherical Fermi surfaces~\cite{Lifshitz-Kaganov:1957}, we extend the field domain in Eqs.~(\ref{Eslab2-short}), (\ref{Eslab2-long}), and (\ref{Eslab2-local}) to $B_0\lesssim B^\star$. The main qualitative difference between the nonlocal and local regimes is that the chiral anomaly enhances the transmission amplitude in some interval of $\omega$ for the former one, while it suppresses the amplitude at any $\omega$ in the local regime. Scaling of the transmission amplitude with the film thickness $L$ at $\xi\ll 1$ is controlled by a single parameter $L/\delta(\omega)$, see Eq.~(\ref{Eslab2-local}). In the nonlocal regime, the dependence on $L$ is defined by the normal-skin and diffusion lengths, $\delta(\omega)$ and $\xi\delta(\omega)$, respectively. In certain intervals of $L$, the anomalous correction competes with the normal-skin term in $E_{\parallel\rm out}$ [see Eqs.~(\ref{Eslab2-short}) and (\ref{Eslab2-long})], resulting in the negative values of $\left|E_{\rm \parallel out}\right|/\left|E_{\rm out}(B_0=0)\right|-1$. However, with the raise of frequency, the anomalous term could win over the normal-skin one, as is illustrated by Fig.~\ref{fig:anisotropy-2}.

{\it Discussion and Summary.---}
We showed that the chiral anomaly may lead to a nonlocal current response of a Weyl or Dirac semimetal even under the conditions of the normal skin effect. The length scale for the nonlocality is determined by the diffusion length of the valley charge imbalance, which does not violate the local electric charge neutrality. This nonlocality is manifested in the penetration and transmission of electromagnetic waves if the diffusion length exceeds the normal-skin depth. Such a regime may be possible in sufficiently clean materials.

The chiral anomaly is activated by a static magnetic field $\mathbf{B}_0$ applied parallel to the surface of the material. The anomaly affects the transmission of an electromagnetic wave with the electric field $\mathbf{E}_\parallel$ parallel to $\mathbf{B}_0$. In this case, the penetration of the field is sensitive to the competition between the normal and anomalous mechanisms of the electromagnetic field propagation in the material. The penetration of the component of the electric field $\mathbf{E}_\perp$ orthogonal to $\mathbf{B}_0$ is unaffected by the anomaly.

We developed a detailed prediction for the field transmission across the film; see Eqs.~(\ref{Eslab2-short}) and (\ref{Eslab2-long}) for films thick and thin compared with the diffusion length, respectively, as well as Eq.~(\ref{Eslab2-local}) for the local response regime. In view of a weaker decay of the anomalous components, it might be possible to achieve an enhancement of the electromagnetic wave penetration depth in the nonlocal regime; see Fig.~\ref{fig:anisotropy-2}. Furthermore, the anomalous part of the transmitted field in the local and nonlocal regimes of the current response is characterized by a different scaling with frequency, {\sl cf.} Eqs.~(\ref{Eslab2-short}) and (\ref{Eslab2-long}) with Eq.~(\ref{Eslab2-local}). These features may allow one to identify the nonlocality, even if the electron transport parameters of a sample are not known in advance.

\begin{acknowledgments}
{\it Acknowledgments.--}
The authors acknowledge useful communications with N.~P.~Armitage, E.~V.~Gorbar, and I.~A.~Shovkovy. This work is supported by NSF Grant No.~DMR-2002275 (L.I.G.). P.O.S. acknowledges support through the Yale Prize Postdoctoral Fellowship in Condensed Matter Theory. This work was performed in part at Aspen Center for Physics, which is supported by NSF Grant No.~PHY-1607611.
\end{acknowledgments}

\vspace{-0.7cm}

\bibliography{library-short}

\begin{thebibliography}{47}%
\makeatletter
\providecommand \@ifxundefined [1]{%
 \@ifx{#1\undefined}
}%
\providecommand \@ifnum [1]{%
 \ifnum #1\expandafter \@firstoftwo
 \else \expandafter \@secondoftwo
 \fi
}%
\providecommand \@ifx [1]{%
 \ifx #1\expandafter \@firstoftwo
 \else \expandafter \@secondoftwo
 \fi
}%
\providecommand \natexlab [1]{#1}%
\providecommand \enquote  [1]{``#1''}%
\providecommand \bibnamefont  [1]{#1}%
\providecommand \bibfnamefont [1]{#1}%
\providecommand \citenamefont [1]{#1}%
\providecommand \href@noop [0]{\@secondoftwo}%
\providecommand \href [0]{\begingroup \@sanitize@url \@href}%
\providecommand \@href[1]{\@@startlink{#1}\@@href}%
\providecommand \@@href[1]{\endgroup#1\@@endlink}%
\providecommand \@sanitize@url [0]{\catcode `\\12\catcode `\$12\catcode
  `\&12\catcode `\#12\catcode `\^12\catcode `\_12\catcode `\%12\relax}%
\providecommand \@@startlink[1]{}%
\providecommand \@@endlink[0]{}%
\providecommand \url  [0]{\begingroup\@sanitize@url \@url }%
\providecommand \@url [1]{\endgroup\@href {#1}{\urlprefix }}%
\providecommand \urlprefix  [0]{URL }%
\providecommand \Eprint [0]{\href }%
\providecommand \doibase [0]{https://doi.org/}%
\providecommand \selectlanguage [0]{\@gobble}%
\providecommand \bibinfo  [0]{\@secondoftwo}%
\providecommand \bibfield  [0]{\@secondoftwo}%
\providecommand \translation [1]{[#1]}%
\providecommand \BibitemOpen [0]{}%
\providecommand \bibitemStop [0]{}%
\providecommand \bibitemNoStop [0]{.\EOS\space}%
\providecommand \EOS [0]{\spacefactor3000\relax}%
\providecommand \BibitemShut  [1]{\csname bibitem#1\endcsname}%
\let\auto@bib@innerbib\@empty
\bibitem [{\citenamefont {Nielsen}\ and\ \citenamefont
  {Ninomiya}(1983)}]{Nielsen:1983}%
  \BibitemOpen
  \bibfield  {author} {\bibinfo {author} {\bibfnamefont {H.}~\bibnamefont
  {Nielsen}}\ and\ \bibinfo {author} {\bibfnamefont {M.}~\bibnamefont
  {Ninomiya}},\ }\bibfield  {title} {\bibinfo {title} {{The Adler-Bell-Jackiw
  anomaly and Weyl fermions in a crystal}},\ }\href
  {https://doi.org/10.1016/0370-2693(83)91529-0} {\bibfield  {journal}
  {\bibinfo  {journal} {Phys. Lett. B}\ }\textbf {\bibinfo {volume} {130}},\
  \bibinfo {pages} {389} (\bibinfo {year} {1983})}\BibitemShut {NoStop}%
\bibitem [{\citenamefont {Adler}(1969)}]{Adler:1969}%
  \BibitemOpen
  \bibfield  {author} {\bibinfo {author} {\bibfnamefont {S.~L.}\ \bibnamefont
  {Adler}},\ }\bibfield  {title} {\bibinfo {title} {{Axial-vector vertex in
  spinor electrodynamics}},\ }\href {https://doi.org/10.1103/PhysRev.177.2426}
  {\bibfield  {journal} {\bibinfo  {journal} {Phys. Rev.}\ }\textbf {\bibinfo
  {volume} {177}},\ \bibinfo {pages} {2426} (\bibinfo {year}
  {1969})}\BibitemShut {NoStop}%
\bibitem [{\citenamefont {Bell}\ and\ \citenamefont {Jackiw}(1969)}]{BJ:1969}%
  \BibitemOpen
  \bibfield  {author} {\bibinfo {author} {\bibfnamefont {J.~S.}\ \bibnamefont
  {Bell}}\ and\ \bibinfo {author} {\bibfnamefont {R.}~\bibnamefont {Jackiw}},\
  }\bibfield  {title} {\bibinfo {title} {{A PCAC puzzle: $\pi_0\to\gamma
  \gamma$ in the $\sigma$-model}},\ }\href {https://doi.org/10.1007/BF02823296}
  {\bibfield  {journal} {\bibinfo  {journal} {Nuovo Cim. A}\ }\textbf {\bibinfo
  {volume} {60}},\ \bibinfo {pages} {47} (\bibinfo {year} {1969})}\BibitemShut
  {NoStop}%
\bibitem [{\citenamefont {Bevan}\ \emph {et~al.}(1997)\citenamefont {Bevan},
  \citenamefont {Manninen}, \citenamefont {Cook}, \citenamefont {Hook},
  \citenamefont {Hall}, \citenamefont {Vachaspati},\ and\ \citenamefont
  {Volovik}}]{Bevan-Volovik:1997}%
  \BibitemOpen
  \bibfield  {author} {\bibinfo {author} {\bibfnamefont {T.~D.~C.}\
  \bibnamefont {Bevan}}, \bibinfo {author} {\bibfnamefont {A.~J.}\ \bibnamefont
  {Manninen}}, \bibinfo {author} {\bibfnamefont {J.~B.}\ \bibnamefont {Cook}},
  \bibinfo {author} {\bibfnamefont {J.~R.}\ \bibnamefont {Hook}}, \bibinfo
  {author} {\bibfnamefont {H.~E.}\ \bibnamefont {Hall}}, \bibinfo {author}
  {\bibfnamefont {T.}~\bibnamefont {Vachaspati}},\ and\ \bibinfo {author}
  {\bibfnamefont {G.~E.}\ \bibnamefont {Volovik}},\ }\bibfield  {title}
  {\bibinfo {title} {{Momentum creation by vortices in superfluid ${}^3$He as a
  model of primordial baryogenesis}},\ }\href
  {https://doi.org/10.1038/386689a0} {\bibfield  {journal} {\bibinfo  {journal}
  {Nature}\ }\textbf {\bibinfo {volume} {386}},\ \bibinfo {pages} {689}
  (\bibinfo {year} {1997})}\BibitemShut {NoStop}%
\bibitem [{\citenamefont {Hasan}\ \emph {et~al.}(2017)\citenamefont {Hasan},
  \citenamefont {Xu}, \citenamefont {Belopolski},\ and\ \citenamefont
  {Huang}}]{Hasan-Huang:rev-2017}%
  \BibitemOpen
  \bibfield  {author} {\bibinfo {author} {\bibfnamefont {M.~Z.}\ \bibnamefont
  {Hasan}}, \bibinfo {author} {\bibfnamefont {S.-Y.}\ \bibnamefont {Xu}},
  \bibinfo {author} {\bibfnamefont {I.}~\bibnamefont {Belopolski}},\ and\
  \bibinfo {author} {\bibfnamefont {S.-M.}\ \bibnamefont {Huang}},\ }\bibfield
  {title} {\bibinfo {title} {{Discovery of Weyl fermion semimetals and
  topological Fermi arc states}},\ }\href
  {https://doi.org/10.1146/annurev-conmatphys-031016-025225} {\bibfield
  {journal} {\bibinfo  {journal} {Annu. Rev. Condens. Matter Phys.}\ }\textbf
  {\bibinfo {volume} {8}},\ \bibinfo {pages} {289} (\bibinfo {year}
  {2017})}\BibitemShut {NoStop}%
\bibitem [{\citenamefont {Armitage}\ \emph {et~al.}(2018)\citenamefont
  {Armitage}, \citenamefont {Mele},\ and\ \citenamefont
  {Vishwanath}}]{Armitage:rev-2018}%
  \BibitemOpen
  \bibfield  {author} {\bibinfo {author} {\bibfnamefont {N.~P.}\ \bibnamefont
  {Armitage}}, \bibinfo {author} {\bibfnamefont {E.~J.}\ \bibnamefont {Mele}},\
  and\ \bibinfo {author} {\bibfnamefont {A.}~\bibnamefont {Vishwanath}},\
  }\bibfield  {title} {\bibinfo {title} {{Weyl and Dirac semimetals in
  three-dimensional solids}},\ }\href
  {https://doi.org/10.1103/RevModPhys.90.015001} {\bibfield  {journal}
  {\bibinfo  {journal} {Rev. Mod. Phys.}\ }\textbf {\bibinfo {volume} {90}},\
  \bibinfo {pages} {015001} (\bibinfo {year} {2018})}\BibitemShut {NoStop}%
\bibitem [{\citenamefont {Gorbar}\ \emph {et~al.}(2021)\citenamefont {Gorbar},
  \citenamefont {Miransky}, \citenamefont {Shovkovy},\ and\ \citenamefont
  {Sukhachov}}]{GMSS:book}%
  \BibitemOpen
  \bibfield  {author} {\bibinfo {author} {\bibfnamefont {E.~V.}\ \bibnamefont
  {Gorbar}}, \bibinfo {author} {\bibfnamefont {V.~A.}\ \bibnamefont
  {Miransky}}, \bibinfo {author} {\bibfnamefont {I.~A.}\ \bibnamefont
  {Shovkovy}},\ and\ \bibinfo {author} {\bibfnamefont {P.~O.}\ \bibnamefont
  {Sukhachov}},\ }\href {https://doi.org/10.1142/11475} {\emph {\bibinfo
  {title} {{Electronic Properties of Dirac and Weyl Semimetals}}}}\ (\bibinfo
  {publisher} {World Scientific},\ \bibinfo {address} {Singapore},\ \bibinfo
  {year} {2021})\BibitemShut {NoStop}%
\bibitem [{\citenamefont {Son}\ and\ \citenamefont
  {Spivak}(2013)}]{Son-Spivak:2013}%
  \BibitemOpen
  \bibfield  {author} {\bibinfo {author} {\bibfnamefont {D.~T.}\ \bibnamefont
  {Son}}\ and\ \bibinfo {author} {\bibfnamefont {B.~Z.}\ \bibnamefont
  {Spivak}},\ }\bibfield  {title} {\bibinfo {title} {{Chiral anomaly and
  classical negative magnetoresistance of Weyl metals}},\ }\href
  {https://doi.org/10.1103/PhysRevB.88.104412} {\bibfield  {journal} {\bibinfo
  {journal} {Phys. Rev. B}\ }\textbf {\bibinfo {volume} {88}},\ \bibinfo
  {pages} {104412} (\bibinfo {year} {2013})}\BibitemShut {NoStop}%
\bibitem [{\citenamefont {Hosur}\ and\ \citenamefont
  {Qi}(2013)}]{Hosur-Qi:rev-2013}%
  \BibitemOpen
  \bibfield  {author} {\bibinfo {author} {\bibfnamefont {P.}~\bibnamefont
  {Hosur}}\ and\ \bibinfo {author} {\bibfnamefont {X.}~\bibnamefont {Qi}},\
  }\bibfield  {title} {\bibinfo {title} {{Recent developments in transport
  phenomena in Weyl semimetals}},\ }\href
  {https://doi.org/10.1016/j.crhy.2013.10.010} {\bibfield  {journal} {\bibinfo
  {journal} {Comptes Rendus Phys.}\ }\textbf {\bibinfo {volume} {14}},\
  \bibinfo {pages} {857} (\bibinfo {year} {2013})}\BibitemShut {NoStop}%
\bibitem [{\citenamefont {Burkov}(2015)}]{Burkov:rev-2015}%
  \BibitemOpen
  \bibfield  {author} {\bibinfo {author} {\bibfnamefont {A.~A.}\ \bibnamefont
  {Burkov}},\ }\bibfield  {title} {\bibinfo {title} {{Chiral anomaly and
  transport in Weyl metals}},\ }\href
  {https://doi.org/10.1088/0953-8984/27/11/113201} {\bibfield  {journal}
  {\bibinfo  {journal} {J. Phys. Condens. Matter}\ }\textbf {\bibinfo {volume}
  {27}},\ \bibinfo {pages} {113201} (\bibinfo {year} {2015})}\BibitemShut
  {NoStop}%
\bibitem [{\citenamefont {Gorbar}\ \emph {et~al.}(2018)\citenamefont {Gorbar},
  \citenamefont {Miransky}, \citenamefont {Shovkovy},\ and\ \citenamefont
  {Sukhachov}}]{Gorbar:2017lnp}%
  \BibitemOpen
  \bibfield  {author} {\bibinfo {author} {\bibfnamefont {E.~V.}\ \bibnamefont
  {Gorbar}}, \bibinfo {author} {\bibfnamefont {V.~A.}\ \bibnamefont
  {Miransky}}, \bibinfo {author} {\bibfnamefont {I.~A.}\ \bibnamefont
  {Shovkovy}},\ and\ \bibinfo {author} {\bibfnamefont {P.~O.}\ \bibnamefont
  {Sukhachov}},\ }\bibfield  {title} {\bibinfo {title} {{Anomalous transport
  properties of Dirac and Weyl semimetals (Review Article)}},\ }\href
  {https://doi.org/10.1063/1.5037551} {\bibfield  {journal} {\bibinfo
  {journal} {Low Temp. Phys.}\ }\textbf {\bibinfo {volume} {44}},\ \bibinfo
  {pages} {487} (\bibinfo {year} {2018})}\BibitemShut {NoStop}%
\bibitem [{\citenamefont {Hu}\ \emph {et~al.}(2019)\citenamefont {Hu},
  \citenamefont {Xu}, \citenamefont {Ni},\ and\ \citenamefont
  {Mao}}]{Hu-Mao:rev-2019}%
  \BibitemOpen
  \bibfield  {author} {\bibinfo {author} {\bibfnamefont {J.}~\bibnamefont
  {Hu}}, \bibinfo {author} {\bibfnamefont {S.-Y.}\ \bibnamefont {Xu}}, \bibinfo
  {author} {\bibfnamefont {N.}~\bibnamefont {Ni}},\ and\ \bibinfo {author}
  {\bibfnamefont {Z.}~\bibnamefont {Mao}},\ }\bibfield  {title} {\bibinfo
  {title} {{Transport of Topological Semimetals}},\ }\href
  {https://doi.org/10.1146/annurev-matsci-070218-010023} {\bibfield  {journal}
  {\bibinfo  {journal} {Annu. Rev. Mater. Res.}\ }\textbf {\bibinfo {volume}
  {49}},\ \bibinfo {pages} {207} (\bibinfo {year} {2019})}\BibitemShut
  {NoStop}%
\bibitem [{\citenamefont {Ong}\ and\ \citenamefont
  {Liang}(2021)}]{Ong-Liang:rev-2020}%
  \BibitemOpen
  \bibfield  {author} {\bibinfo {author} {\bibfnamefont {N.~P.}\ \bibnamefont
  {Ong}}\ and\ \bibinfo {author} {\bibfnamefont {S.}~\bibnamefont {Liang}},\
  }\bibfield  {title} {\bibinfo {title} {{Experimental signatures of the chiral
  anomaly in Dirac–Weyl semimetals}},\ }\href
  {https://doi.org/10.1038/s42254-021-00310-9} {\bibfield  {journal} {\bibinfo
  {journal} {Nat. Rev. Phys.}\ }\textbf {\bibinfo {volume} {3}},\ \bibinfo
  {pages} {394} (\bibinfo {year} {2021})}\BibitemShut {NoStop}%
\bibitem [{\citenamefont {dos Reis}\ \emph {et~al.}(2016)\citenamefont {dos
  Reis}, \citenamefont {Ajeesh}, \citenamefont {Kumar}, \citenamefont {Arnold},
  \citenamefont {Shekhar}, \citenamefont {Naumann}, \citenamefont {Schmidt},
  \citenamefont {Nicklas},\ and\ \citenamefont
  {Hassinger}}]{Reis-Hassinger:2016}%
  \BibitemOpen
  \bibfield  {author} {\bibinfo {author} {\bibfnamefont {R.~D.}\ \bibnamefont
  {dos Reis}}, \bibinfo {author} {\bibfnamefont {M.~O.}\ \bibnamefont
  {Ajeesh}}, \bibinfo {author} {\bibfnamefont {N.}~\bibnamefont {Kumar}},
  \bibinfo {author} {\bibfnamefont {F.}~\bibnamefont {Arnold}}, \bibinfo
  {author} {\bibfnamefont {C.}~\bibnamefont {Shekhar}}, \bibinfo {author}
  {\bibfnamefont {M.}~\bibnamefont {Naumann}}, \bibinfo {author} {\bibfnamefont
  {M.}~\bibnamefont {Schmidt}}, \bibinfo {author} {\bibfnamefont
  {M.}~\bibnamefont {Nicklas}},\ and\ \bibinfo {author} {\bibfnamefont
  {E.}~\bibnamefont {Hassinger}},\ }\bibfield  {title} {\bibinfo {title} {{On
  the search for the chiral anomaly in Weyl semimetals: the negative
  longitudinal magnetoresistance}},\ }\href
  {https://doi.org/10.1088/1367-2630/18/8/085006} {\bibfield  {journal}
  {\bibinfo  {journal} {New J. Phys.}\ }\textbf {\bibinfo {volume} {18}},\
  \bibinfo {pages} {085006} (\bibinfo {year} {2016})}\BibitemShut {NoStop}%
\bibitem [{\citenamefont {Liang}\ \emph {et~al.}(2018)\citenamefont {Liang},
  \citenamefont {Lin}, \citenamefont {Kushwaha}, \citenamefont {Xing},
  \citenamefont {Ni}, \citenamefont {Cava},\ and\ \citenamefont
  {Ong}}]{Liang-Ong:2018}%
  \BibitemOpen
  \bibfield  {author} {\bibinfo {author} {\bibfnamefont {S.}~\bibnamefont
  {Liang}}, \bibinfo {author} {\bibfnamefont {J.}~\bibnamefont {Lin}}, \bibinfo
  {author} {\bibfnamefont {S.}~\bibnamefont {Kushwaha}}, \bibinfo {author}
  {\bibfnamefont {J.}~\bibnamefont {Xing}}, \bibinfo {author} {\bibfnamefont
  {N.}~\bibnamefont {Ni}}, \bibinfo {author} {\bibfnamefont {R.~J.}\
  \bibnamefont {Cava}},\ and\ \bibinfo {author} {\bibfnamefont {N.~P.}\
  \bibnamefont {Ong}},\ }\bibfield  {title} {\bibinfo {title} {{Experimental
  tests of the chiral anomaly magnetoresistance in the Dirac-Weyl semimetals
  Na$_3$Bi and GdPtBi}},\ }\href {https://doi.org/10.1103/PhysRevX.8.031002}
  {\bibfield  {journal} {\bibinfo  {journal} {Phys. Rev. X}\ }\textbf {\bibinfo
  {volume} {8}},\ \bibinfo {pages} {031002} (\bibinfo {year}
  {2018})}\BibitemShut {NoStop}%
\bibitem [{\citenamefont {Goswami}\ \emph {et~al.}(2015)\citenamefont
  {Goswami}, \citenamefont {Pixley},\ and\ \citenamefont {{Das
  Sarma}}}]{Goswami-DasSarma:2015}%
  \BibitemOpen
  \bibfield  {author} {\bibinfo {author} {\bibfnamefont {P.}~\bibnamefont
  {Goswami}}, \bibinfo {author} {\bibfnamefont {J.~H.}\ \bibnamefont
  {Pixley}},\ and\ \bibinfo {author} {\bibfnamefont {S.}~\bibnamefont {{Das
  Sarma}}},\ }\bibfield  {title} {\bibinfo {title} {{Axial anomaly and
  longitudinal magnetoresistance of a generic three-dimensional metal}},\
  }\href {https://doi.org/10.1103/PhysRevB.92.075205} {\bibfield  {journal}
  {\bibinfo  {journal} {Phys. Rev. B}\ }\textbf {\bibinfo {volume} {92}},\
  \bibinfo {pages} {075205} (\bibinfo {year} {2015})}\BibitemShut {NoStop}%
\bibitem [{\citenamefont {Burkov}(2018)}]{Burkov:2018-OptCond}%
  \BibitemOpen
  \bibfield  {author} {\bibinfo {author} {\bibfnamefont {A.~A.}\ \bibnamefont
  {Burkov}},\ }\bibfield  {title} {\bibinfo {title} {{Dynamical density
  response and optical conductivity in topological metals}},\ }\href
  {https://doi.org/10.1103/PhysRevB.98.165123} {\bibfield  {journal} {\bibinfo
  {journal} {Phys. Rev. B}\ }\textbf {\bibinfo {volume} {98}},\ \bibinfo
  {pages} {165123} (\bibinfo {year} {2018})}\BibitemShut {NoStop}%
\bibitem [{\citenamefont {Cheng}\ \emph {et~al.}(2021)\citenamefont {Cheng},
  \citenamefont {Schumann}, \citenamefont {Stemmer},\ and\ \citenamefont
  {Armitage}}]{Cheng-Armitage:2019}%
  \BibitemOpen
  \bibfield  {author} {\bibinfo {author} {\bibfnamefont {B.}~\bibnamefont
  {Cheng}}, \bibinfo {author} {\bibfnamefont {T.}~\bibnamefont {Schumann}},
  \bibinfo {author} {\bibfnamefont {S.}~\bibnamefont {Stemmer}},\ and\ \bibinfo
  {author} {\bibfnamefont {N.~P.}\ \bibnamefont {Armitage}},\ }\bibfield
  {title} {\bibinfo {title} {{Probing charge pumping and relaxation of the
  chiral anomaly in a Dirac semimetal}},\ }\href
  {https://doi.org/10.1126/sciadv.abg0914} {\bibfield  {journal} {\bibinfo
  {journal} {Sci. Adv.}\ }\textbf {\bibinfo {volume} {7}},\ \bibinfo {pages}
  {eabg0914} (\bibinfo {year} {2021})}\BibitemShut {NoStop}%
\bibitem [{\citenamefont {Abrikosov}(2017)}]{Abrikosov:book-1988}%
  \BibitemOpen
  \bibfield  {author} {\bibinfo {author} {\bibfnamefont {A.~A.}\ \bibnamefont
  {Abrikosov}},\ }\href {https://books.google.ca/books?id=tTo2DwAAQBAJ} {\emph
  {\bibinfo {title} {{Fundamentals of the Theory of Metals}}}}\ (\bibinfo
  {publisher} {Courier Dover Publications},\ \bibinfo {address} {New York},\
  \bibinfo {year} {2017})\BibitemShut {NoStop}%
\bibitem [{\citenamefont {Vilenkin}(1980)}]{Vilenkin:1980}%
  \BibitemOpen
  \bibfield  {author} {\bibinfo {author} {\bibfnamefont {A.}~\bibnamefont
  {Vilenkin}},\ }\bibfield  {title} {\bibinfo {title} {{Equilibrium
  parity-violating current in a magnetic field}},\ }\href
  {https://doi.org/10.1103/PhysRevD.22.3080} {\bibfield  {journal} {\bibinfo
  {journal} {Phys. Rev. D}\ }\textbf {\bibinfo {volume} {22}},\ \bibinfo
  {pages} {3080} (\bibinfo {year} {1980})}\BibitemShut {NoStop}%
\bibitem [{\citenamefont {Fukushima}\ \emph {et~al.}(2008)\citenamefont
  {Fukushima}, \citenamefont {Kharzeev},\ and\ \citenamefont
  {Warringa}}]{Fukushima:2008}%
  \BibitemOpen
  \bibfield  {author} {\bibinfo {author} {\bibfnamefont {K.}~\bibnamefont
  {Fukushima}}, \bibinfo {author} {\bibfnamefont {D.~E.}\ \bibnamefont
  {Kharzeev}},\ and\ \bibinfo {author} {\bibfnamefont {H.~J.}\ \bibnamefont
  {Warringa}},\ }\bibfield  {title} {\bibinfo {title} {{Chiral magnetic
  effect}},\ }\href {https://doi.org/10.1103/PhysRevD.78.074033} {\bibfield
  {journal} {\bibinfo  {journal} {Phys. Rev. D}\ }\textbf {\bibinfo {volume}
  {78}},\ \bibinfo {pages} {074033} (\bibinfo {year} {2008})}\BibitemShut
  {NoStop}%
\bibitem [{\citenamefont {Parameswaran}\ \emph {et~al.}(2014)\citenamefont
  {Parameswaran}, \citenamefont {Grover}, \citenamefont {Abanin}, \citenamefont
  {Pesin},\ and\ \citenamefont {Vishwanath}}]{Parameswaran-Vishwanath:2014}%
  \BibitemOpen
  \bibfield  {author} {\bibinfo {author} {\bibfnamefont {S.~A.}\ \bibnamefont
  {Parameswaran}}, \bibinfo {author} {\bibfnamefont {T.}~\bibnamefont
  {Grover}}, \bibinfo {author} {\bibfnamefont {D.~A.}\ \bibnamefont {Abanin}},
  \bibinfo {author} {\bibfnamefont {D.~A.}\ \bibnamefont {Pesin}},\ and\
  \bibinfo {author} {\bibfnamefont {A.}~\bibnamefont {Vishwanath}},\ }\bibfield
   {title} {\bibinfo {title} {{Probing the chiral anomaly with nonlocal
  transport in three-dimensional topological semimetals}},\ }\href
  {https://doi.org/10.1103/PhysRevX.4.031035} {\bibfield  {journal} {\bibinfo
  {journal} {Phys. Rev. X}\ }\textbf {\bibinfo {volume} {4}},\ \bibinfo {pages}
  {031035} (\bibinfo {year} {2014})}\BibitemShut {NoStop}%
\bibitem [{\citenamefont {Zhang}\ \emph {et~al.}(2017)\citenamefont {Zhang},
  \citenamefont {Zhang}, \citenamefont {Wang}, \citenamefont {Liu},
  \citenamefont {Chen}, \citenamefont {Lu}, \citenamefont {Liang},
  \citenamefont {Cao}, \citenamefont {Yuan}, \citenamefont {Tang},
  \citenamefont {Li}, \citenamefont {Zhou}, \citenamefont {Gu}, \citenamefont
  {Wu}, \citenamefont {Zou},\ and\ \citenamefont {Xiu}}]{Zhang-Xiu:2017}%
  \BibitemOpen
  \bibfield  {author} {\bibinfo {author} {\bibfnamefont {C.}~\bibnamefont
  {Zhang}}, \bibinfo {author} {\bibfnamefont {E.}~\bibnamefont {Zhang}},
  \bibinfo {author} {\bibfnamefont {W.}~\bibnamefont {Wang}}, \bibinfo {author}
  {\bibfnamefont {Y.}~\bibnamefont {Liu}}, \bibinfo {author} {\bibfnamefont
  {Z.-G.}\ \bibnamefont {Chen}}, \bibinfo {author} {\bibfnamefont
  {S.}~\bibnamefont {Lu}}, \bibinfo {author} {\bibfnamefont {S.}~\bibnamefont
  {Liang}}, \bibinfo {author} {\bibfnamefont {J.}~\bibnamefont {Cao}}, \bibinfo
  {author} {\bibfnamefont {X.}~\bibnamefont {Yuan}}, \bibinfo {author}
  {\bibfnamefont {L.}~\bibnamefont {Tang}}, \bibinfo {author} {\bibfnamefont
  {Q.}~\bibnamefont {Li}}, \bibinfo {author} {\bibfnamefont {C.}~\bibnamefont
  {Zhou}}, \bibinfo {author} {\bibfnamefont {T.}~\bibnamefont {Gu}}, \bibinfo
  {author} {\bibfnamefont {Y.}~\bibnamefont {Wu}}, \bibinfo {author}
  {\bibfnamefont {J.}~\bibnamefont {Zou}},\ and\ \bibinfo {author}
  {\bibfnamefont {F.}~\bibnamefont {Xiu}},\ }\bibfield  {title} {\bibinfo
  {title} {{Room-temperature chiral charge pumping in Dirac semimetals}},\
  }\href {https://doi.org/10.1038/ncomms13741} {\bibfield  {journal} {\bibinfo
  {journal} {Nat. Commun.}\ }\textbf {\bibinfo {volume} {8}},\ \bibinfo {pages}
  {13741} (\bibinfo {year} {2017})}\BibitemShut {NoStop}%
\bibitem [{\citenamefont {de~Boer}\ \emph {et~al.}(2019)\citenamefont
  {de~Boer}, \citenamefont {Wielens}, \citenamefont {Voerman}, \citenamefont
  {de~Ronde}, \citenamefont {Huang}, \citenamefont {Golden}, \citenamefont
  {Li},\ and\ \citenamefont {Brinkman}}]{Boer-Brinkman:2019}%
  \BibitemOpen
  \bibfield  {author} {\bibinfo {author} {\bibfnamefont {J.~C.}\ \bibnamefont
  {de~Boer}}, \bibinfo {author} {\bibfnamefont {D.~H.}\ \bibnamefont
  {Wielens}}, \bibinfo {author} {\bibfnamefont {J.~A.}\ \bibnamefont
  {Voerman}}, \bibinfo {author} {\bibfnamefont {B.}~\bibnamefont {de~Ronde}},
  \bibinfo {author} {\bibfnamefont {Y.}~\bibnamefont {Huang}}, \bibinfo
  {author} {\bibfnamefont {M.~S.}\ \bibnamefont {Golden}}, \bibinfo {author}
  {\bibfnamefont {C.}~\bibnamefont {Li}},\ and\ \bibinfo {author}
  {\bibfnamefont {A.}~\bibnamefont {Brinkman}},\ }\bibfield  {title} {\bibinfo
  {title} {{Nonlocal signatures of the chiral magnetic effect in the Dirac
  semimetal Bi$_{0.97}$Sb$_{0.03}$}},\ }\href
  {https://doi.org/10.1103/PhysRevB.99.085124} {\bibfield  {journal} {\bibinfo
  {journal} {Phys. Rev. B}\ }\textbf {\bibinfo {volume} {99}},\ \bibinfo
  {pages} {085124} (\bibinfo {year} {2019})}\BibitemShut {NoStop}%
\bibitem [{Note1()}]{Note1}%
  \BibitemOpen
  \bibinfo {note} {The main qualitative results of our study should hold for a
  time-reversal symmetry broken Weyl semimetal as well. However, such systems
  have other phenomena that also affect the propagation and reflection of
  electromagnetic waves, {\protect \sl e.g.}, anomalous Kerr and Faraday
  effects~\cite {Kargarian-Trivedi:2015}. These phenomena are ignored in this
  work where we focus on the effects of the chiral anomaly.}\BibitemShut
  {Stop}%
\bibitem [{\citenamefont {Jadidi}\ \emph {et~al.}(2020)\citenamefont {Jadidi},
  \citenamefont {Kargarian}, \citenamefont {Mittendorff}, \citenamefont
  {Aytac}, \citenamefont {Shen}, \citenamefont {K{\"{o}}nig-Otto},
  \citenamefont {Winnerl}, \citenamefont {Ni}, \citenamefont {Gaeta},
  \citenamefont {Murphy},\ and\ \citenamefont {Drew}}]{Jadidi-Drew-TaAs:2019}%
  \BibitemOpen
  \bibfield  {author} {\bibinfo {author} {\bibfnamefont {M.~M.}\ \bibnamefont
  {Jadidi}}, \bibinfo {author} {\bibfnamefont {M.}~\bibnamefont {Kargarian}},
  \bibinfo {author} {\bibfnamefont {M.}~\bibnamefont {Mittendorff}}, \bibinfo
  {author} {\bibfnamefont {Y.}~\bibnamefont {Aytac}}, \bibinfo {author}
  {\bibfnamefont {B.}~\bibnamefont {Shen}}, \bibinfo {author} {\bibfnamefont
  {J.~C.}\ \bibnamefont {K{\"{o}}nig-Otto}}, \bibinfo {author} {\bibfnamefont
  {S.}~\bibnamefont {Winnerl}}, \bibinfo {author} {\bibfnamefont
  {N.}~\bibnamefont {Ni}}, \bibinfo {author} {\bibfnamefont {A.~L.}\
  \bibnamefont {Gaeta}}, \bibinfo {author} {\bibfnamefont {T.~E.}\ \bibnamefont
  {Murphy}},\ and\ \bibinfo {author} {\bibfnamefont {H.~D.}\ \bibnamefont
  {Drew}},\ }\bibfield  {title} {\bibinfo {title} {{Nonlinear optical control
  of chiral charge pumping in a topological Weyl semimetal}},\ }\href
  {https://doi.org/10.1103/PhysRevB.102.245123} {\bibfield  {journal} {\bibinfo
   {journal} {Phys. Rev. B}\ }\textbf {\bibinfo {volume} {102}},\ \bibinfo
  {pages} {245123} (\bibinfo {year} {2020})}\BibitemShut {NoStop}%
\bibitem [{Note2()}]{Note2}%
  \BibitemOpen
  \bibinfo {note} {The cyclotron frequency for the classically-weak magnetic
  field is smaller than the intra-node scattering rate, {\protect \sl i.e.},
  $\omega _c\ll \unhbox \voidb@x \hbox {min}\left \protect \{1/\tau _{\alpha
  ,\alpha }\right \protect \}$, where $1/\tau _{\alpha ,\alpha }$ is the
  intranode scattering rate for Weyl node $\alpha $.}\BibitemShut {Stop}%
\bibitem [{Note3()}]{Note3}%
  \BibitemOpen
  \bibinfo {note} {We disregard the contribution of the Fermi arc surface
  states in the current. As is estimated in Ref.~\cite {Chen-Belyanin:2019},
  the relative contribution of the Fermi arcs to the surface impedance is
  negligible if the frequency of the impinging wave is much smaller than the
  plasmon resonance frequency, which is indeed the case in our
  study.}\BibitemShut {Stop}%
\bibitem [{SM()}]{SM}%
  \BibitemOpen
  \href@noop {} {\bibinfo  {journal} {See Supplemental Material for the
  derivations of the kinetic equations and the transmission of electromagnetic
  waves in the nonlocal and local current response regimes. The Supplemental
  Material contains
  Refs.~\cite{Xiao-Niu:rev-2010,Son:2013,Stephanov:2012,Landau:t8}}\
  }\BibitemShut {NoStop}%
\bibitem [{\citenamefont {Burkov}(2014)}]{Burkov:2014b}%
  \BibitemOpen
\bibfield  {journal} {  }\bibfield  {author} {\bibinfo {author} {\bibfnamefont
  {A.~A.}\ \bibnamefont {Burkov}},\ }\bibfield  {title} {\bibinfo {title}
  {{Chiral anomaly and diffusive magnetotransport in Weyl metals}},\ }\href
  {https://doi.org/10.1103/PhysRevLett.113.247203} {\bibfield  {journal}
  {\bibinfo  {journal} {Phys. Rev. Lett.}\ }\textbf {\bibinfo {volume} {113}},\
  \bibinfo {pages} {247203} (\bibinfo {year} {2014})}\BibitemShut {NoStop}%
\bibitem [{Note4()}]{Note4}%
  \BibitemOpen
  \bibinfo {note} {The continuity of the derivatives of the tangential
  components of electric fields at the surfaces follows from the continuity of
  the tangential components of the magnetic fields.}\BibitemShut {Stop}%
\bibitem [{Note5()}]{Note5}%
  \BibitemOpen
  \bibinfo {note} {While we use the parameters of the Weyl semimetal TaAs,
  other materials with a simpler band structure might be used to observe the
  proposed anomalous nonlocal effect. For example, we mention the Dirac
  semimetal Cd$_3$As$_2$~\cite
  {Borisenko:2014,Liu-Chen-Cd3As2:2014,Neupane-Hasan-Cd3As2:2014} and the Weyl
  semimetal EuCd$_2$As$_2$~\cite
  {Wang-Canfield:2019,Soh-Boothroyd:2019,Ma-Shi:2019}. Compared to TaAs, they
  have a simpler band structure with only two Dirac points and Weyl nodes,
  respectively.}\BibitemShut {Stop}%
\bibitem [{\citenamefont {Arnold}\ \emph {et~al.}(2016)\citenamefont {Arnold},
  \citenamefont {Naumann}, \citenamefont {Wu}, \citenamefont {Sun},
  \citenamefont {Schmidt}, \citenamefont {Borrmann}, \citenamefont {Felser},
  \citenamefont {Yan},\ and\ \citenamefont {Hassinger}}]{Arnold-Felser:2016b}%
  \BibitemOpen
  \bibfield  {author} {\bibinfo {author} {\bibfnamefont {F.}~\bibnamefont
  {Arnold}}, \bibinfo {author} {\bibfnamefont {M.}~\bibnamefont {Naumann}},
  \bibinfo {author} {\bibfnamefont {S.-C.}\ \bibnamefont {Wu}}, \bibinfo
  {author} {\bibfnamefont {Y.}~\bibnamefont {Sun}}, \bibinfo {author}
  {\bibfnamefont {M.}~\bibnamefont {Schmidt}}, \bibinfo {author} {\bibfnamefont
  {H.}~\bibnamefont {Borrmann}}, \bibinfo {author} {\bibfnamefont
  {C.}~\bibnamefont {Felser}}, \bibinfo {author} {\bibfnamefont
  {B.}~\bibnamefont {Yan}},\ and\ \bibinfo {author} {\bibfnamefont
  {E.}~\bibnamefont {Hassinger}},\ }\bibfield  {title} {\bibinfo {title}
  {{Chiral Weyl pockets and Fermi surface topology of the Weyl semimetal
  TaAs}},\ }\href {https://doi.org/10.1103/PhysRevLett.117.146401} {\bibfield
  {journal} {\bibinfo  {journal} {Phys. Rev. Lett.}\ }\textbf {\bibinfo
  {volume} {117}},\ \bibinfo {pages} {146401} (\bibinfo {year}
  {2016})}\BibitemShut {NoStop}%
\bibitem [{\citenamefont {Zhang}\ \emph {et~al.}(2016)\citenamefont {Zhang},
  \citenamefont {Xu}, \citenamefont {Belopolski}, \citenamefont {Yuan},
  \citenamefont {Lin}, \citenamefont {Tong}, \citenamefont {Bian},
  \citenamefont {Alidoust}, \citenamefont {Lee}, \citenamefont {Huang},
  \citenamefont {Chang}, \citenamefont {Chang}, \citenamefont {Hsu},
  \citenamefont {Jeng}, \citenamefont {Neupane}, \citenamefont {Sanchez},
  \citenamefont {Zheng}, \citenamefont {Wang}, \citenamefont {Lin},
  \citenamefont {Zhang}, \citenamefont {Lu}, \citenamefont {Shen},
  \citenamefont {Neupert}, \citenamefont {{Zahid Hasan}},\ and\ \citenamefont
  {Jia}}]{Zhang-Hasan-TaAs:2016}%
  \BibitemOpen
  \bibfield  {author} {\bibinfo {author} {\bibfnamefont {C.-L.}\ \bibnamefont
  {Zhang}}, \bibinfo {author} {\bibfnamefont {S.-Y.}\ \bibnamefont {Xu}},
  \bibinfo {author} {\bibfnamefont {I.}~\bibnamefont {Belopolski}}, \bibinfo
  {author} {\bibfnamefont {Z.}~\bibnamefont {Yuan}}, \bibinfo {author}
  {\bibfnamefont {Z.}~\bibnamefont {Lin}}, \bibinfo {author} {\bibfnamefont
  {B.}~\bibnamefont {Tong}}, \bibinfo {author} {\bibfnamefont {G.}~\bibnamefont
  {Bian}}, \bibinfo {author} {\bibfnamefont {N.}~\bibnamefont {Alidoust}},
  \bibinfo {author} {\bibfnamefont {C.-C.}\ \bibnamefont {Lee}}, \bibinfo
  {author} {\bibfnamefont {S.-M.}\ \bibnamefont {Huang}}, \bibinfo {author}
  {\bibfnamefont {T.-R.}\ \bibnamefont {Chang}}, \bibinfo {author}
  {\bibfnamefont {G.}~\bibnamefont {Chang}}, \bibinfo {author} {\bibfnamefont
  {C.-H.}\ \bibnamefont {Hsu}}, \bibinfo {author} {\bibfnamefont {H.-T.}\
  \bibnamefont {Jeng}}, \bibinfo {author} {\bibfnamefont {M.}~\bibnamefont
  {Neupane}}, \bibinfo {author} {\bibfnamefont {D.~S.}\ \bibnamefont
  {Sanchez}}, \bibinfo {author} {\bibfnamefont {H.}~\bibnamefont {Zheng}},
  \bibinfo {author} {\bibfnamefont {J.}~\bibnamefont {Wang}}, \bibinfo {author}
  {\bibfnamefont {H.}~\bibnamefont {Lin}}, \bibinfo {author} {\bibfnamefont
  {C.}~\bibnamefont {Zhang}}, \bibinfo {author} {\bibfnamefont {H.-Z.}\
  \bibnamefont {Lu}}, \bibinfo {author} {\bibfnamefont {S.-Q.}\ \bibnamefont
  {Shen}}, \bibinfo {author} {\bibfnamefont {T.}~\bibnamefont {Neupert}},
  \bibinfo {author} {\bibfnamefont {M.}~\bibnamefont {{Zahid Hasan}}},\ and\
  \bibinfo {author} {\bibfnamefont {S.}~\bibnamefont {Jia}},\ }\bibfield
  {title} {\bibinfo {title} {{Signatures of the Adler–Bell–Jackiw chiral
  anomaly in a Weyl fermion semimetal}},\ }\href
  {https://doi.org/10.1038/ncomms10735} {\bibfield  {journal} {\bibinfo
  {journal} {Nat. Commun.}\ }\textbf {\bibinfo {volume} {7}},\ \bibinfo {pages}
  {10735} (\bibinfo {year} {2016})}\BibitemShut {NoStop}%
\bibitem [{\citenamefont {Lifshitz}\ \emph {et~al.}(1957)\citenamefont
  {Lifshitz}, \citenamefont {Azbel},\ and\ \citenamefont
  {Kaganov}}]{Lifshitz-Kaganov:1957}%
  \BibitemOpen
  \bibfield  {author} {\bibinfo {author} {\bibfnamefont {I.~M.}\ \bibnamefont
  {Lifshitz}}, \bibinfo {author} {\bibfnamefont {M.~I.}\ \bibnamefont
  {Azbel}},\ and\ \bibinfo {author} {\bibfnamefont {M.~I.}\ \bibnamefont
  {Kaganov}},\ }\bibfield  {title} {\bibinfo {title} {{The theory of
  galvanomagnetic effects in metals}},\ }\href
  {http://www.jetp.ac.ru/cgi-bin/dn/e_004_01_0041.pdf} {\bibfield  {journal}
  {\bibinfo  {journal} {JETP}\ }\textbf {\bibinfo {volume} {4}},\ \bibinfo
  {pages} {41} (\bibinfo {year} {1957})}\BibitemShut {NoStop}%
\bibitem [{\citenamefont {Kargarian}\ \emph {et~al.}(2015)\citenamefont
  {Kargarian}, \citenamefont {Randeria},\ and\ \citenamefont
  {Trivedi}}]{Kargarian-Trivedi:2015}%
  \BibitemOpen
  \bibfield  {author} {\bibinfo {author} {\bibfnamefont {M.}~\bibnamefont
  {Kargarian}}, \bibinfo {author} {\bibfnamefont {M.}~\bibnamefont
  {Randeria}},\ and\ \bibinfo {author} {\bibfnamefont {N.}~\bibnamefont
  {Trivedi}},\ }\bibfield  {title} {\bibinfo {title} {{Theory of Kerr and
  Faraday rotations and linear dichroism in Topological Weyl Semimetals}},\
  }\href {https://doi.org/10.1038/srep12683} {\bibfield  {journal} {\bibinfo
  {journal} {Sci. Rep.}\ }\textbf {\bibinfo {volume} {5}},\ \bibinfo {pages}
  {12683} (\bibinfo {year} {2015})}\BibitemShut {NoStop}%
\bibitem [{\citenamefont {Chen}\ \emph {et~al.}(2019)\citenamefont {Chen},
  \citenamefont {Kutayiah}, \citenamefont {Oladyshkin}, \citenamefont
  {Tokman},\ and\ \citenamefont {Belyanin}}]{Chen-Belyanin:2019}%
  \BibitemOpen
  \bibfield  {author} {\bibinfo {author} {\bibfnamefont {Q.}~\bibnamefont
  {Chen}}, \bibinfo {author} {\bibfnamefont {A.~R.}\ \bibnamefont {Kutayiah}},
  \bibinfo {author} {\bibfnamefont {I.}~\bibnamefont {Oladyshkin}}, \bibinfo
  {author} {\bibfnamefont {M.}~\bibnamefont {Tokman}},\ and\ \bibinfo {author}
  {\bibfnamefont {A.}~\bibnamefont {Belyanin}},\ }\bibfield  {title} {\bibinfo
  {title} {{Optical properties and electromagnetic modes of Weyl semimetals}},\
  }\href {https://doi.org/10.1103/PhysRevB.99.075137} {\bibfield  {journal}
  {\bibinfo  {journal} {Phys. Rev. B}\ }\textbf {\bibinfo {volume} {99}},\
  \bibinfo {pages} {075137} (\bibinfo {year} {2019})}\BibitemShut {NoStop}%
\bibitem [{\citenamefont {Xiao}\ \emph {et~al.}(2010)\citenamefont {Xiao},
  \citenamefont {Chang},\ and\ \citenamefont {Niu}}]{Xiao-Niu:rev-2010}%
  \BibitemOpen
  \bibfield  {author} {\bibinfo {author} {\bibfnamefont {D.}~\bibnamefont
  {Xiao}}, \bibinfo {author} {\bibfnamefont {M.~C.}\ \bibnamefont {Chang}},\
  and\ \bibinfo {author} {\bibfnamefont {Q.}~\bibnamefont {Niu}},\ }\bibfield
  {title} {\bibinfo {title} {{Berry phase effects on electronic properties}},\
  }\href {https://doi.org/10.1103/RevModPhys.82.1959} {\bibfield  {journal}
  {\bibinfo  {journal} {Rev. Mod. Phys.}\ }\textbf {\bibinfo {volume} {82}},\
  \bibinfo {pages} {1959} (\bibinfo {year} {2010})}\BibitemShut {NoStop}%
\bibitem [{\citenamefont {Son}\ and\ \citenamefont
  {Yamamoto}(2013)}]{Son:2013}%
  \BibitemOpen
  \bibfield  {author} {\bibinfo {author} {\bibfnamefont {D.~T.}\ \bibnamefont
  {Son}}\ and\ \bibinfo {author} {\bibfnamefont {N.}~\bibnamefont {Yamamoto}},\
  }\bibfield  {title} {\bibinfo {title} {{Kinetic theory with Berry curvature
  from quantum field theories}},\ }\href
  {https://doi.org/10.1103/PhysRevD.87.085016} {\bibfield  {journal} {\bibinfo
  {journal} {Phys. Rev. D}\ }\textbf {\bibinfo {volume} {87}},\ \bibinfo
  {pages} {085016} (\bibinfo {year} {2013})}\BibitemShut {NoStop}%
\bibitem [{\citenamefont {Stephanov}\ and\ \citenamefont
  {Yin}(2012)}]{Stephanov:2012}%
  \BibitemOpen
  \bibfield  {author} {\bibinfo {author} {\bibfnamefont {M.~A.}\ \bibnamefont
  {Stephanov}}\ and\ \bibinfo {author} {\bibfnamefont {Y.}~\bibnamefont
  {Yin}},\ }\bibfield  {title} {\bibinfo {title} {{Chiral kinetic theory}},\
  }\href {https://doi.org/10.1103/PhysRevLett.109.162001} {\bibfield  {journal}
  {\bibinfo  {journal} {Phys. Rev. Lett.}\ }\textbf {\bibinfo {volume} {109}},\
  \bibinfo {pages} {162001} (\bibinfo {year} {2012})}\BibitemShut {NoStop}%
\bibitem [{\citenamefont {Landau}\ \emph {et~al.}(1984)\citenamefont {Landau},
  \citenamefont {Lifshits},\ and\ \citenamefont {Pitaevskii}}]{Landau:t8}%
  \BibitemOpen
  \bibfield  {author} {\bibinfo {author} {\bibfnamefont {L.~D.}\ \bibnamefont
  {Landau}}, \bibinfo {author} {\bibfnamefont {E.~M.}\ \bibnamefont
  {Lifshits}},\ and\ \bibinfo {author} {\bibfnamefont {L.~P.}\ \bibnamefont
  {Pitaevskii}},\ }\href
  {https://www.elsevier.com/books/electrodynamics-of-continuous-media/landau/978-0-08-057060-0}
  {\emph {\bibinfo {title} {{Electrodynamics of Continuous Media}}}}\ (\bibinfo
   {publisher} {Butterworth-Heinemann},\ \bibinfo {address} {Oxford},\ \bibinfo
  {year} {1984})\BibitemShut {NoStop}%
\bibitem [{\citenamefont {Borisenko}\ \emph {et~al.}(2014)\citenamefont
  {Borisenko}, \citenamefont {Gibson}, \citenamefont {Evtushinsky},
  \citenamefont {Zabolotnyy}, \citenamefont {B{\"{u}}chner},\ and\
  \citenamefont {Cava}}]{Borisenko:2014}%
  \BibitemOpen
  \bibfield  {author} {\bibinfo {author} {\bibfnamefont {S.}~\bibnamefont
  {Borisenko}}, \bibinfo {author} {\bibfnamefont {Q.}~\bibnamefont {Gibson}},
  \bibinfo {author} {\bibfnamefont {D.}~\bibnamefont {Evtushinsky}}, \bibinfo
  {author} {\bibfnamefont {V.}~\bibnamefont {Zabolotnyy}}, \bibinfo {author}
  {\bibfnamefont {B.}~\bibnamefont {B{\"{u}}chner}},\ and\ \bibinfo {author}
  {\bibfnamefont {R.~J.}\ \bibnamefont {Cava}},\ }\bibfield  {title} {\bibinfo
  {title} {{Experimental realization of a three-dimensional Dirac semimetal}},\
  }\href {https://doi.org/10.1103/PhysRevLett.113.027603} {\bibfield  {journal}
  {\bibinfo  {journal} {Phys. Rev. Lett.}\ }\textbf {\bibinfo {volume} {113}},\
  \bibinfo {pages} {027603} (\bibinfo {year} {2014})}\BibitemShut {NoStop}%
\bibitem [{\citenamefont {Liu}\ \emph {et~al.}(2014)\citenamefont {Liu},
  \citenamefont {Jiang}, \citenamefont {Zhou}, \citenamefont {Wang},
  \citenamefont {Zhang}, \citenamefont {Weng}, \citenamefont {Prabhakaran},
  \citenamefont {Mo}, \citenamefont {Peng}, \citenamefont {Dudin},
  \citenamefont {Kim}, \citenamefont {Hoesch}, \citenamefont {Fang},
  \citenamefont {Dai}, \citenamefont {Shen}, \citenamefont {Feng},
  \citenamefont {Hussain},\ and\ \citenamefont {Chen}}]{Liu-Chen-Cd3As2:2014}%
  \BibitemOpen
  \bibfield  {author} {\bibinfo {author} {\bibfnamefont {Z.~K.}\ \bibnamefont
  {Liu}}, \bibinfo {author} {\bibfnamefont {J.}~\bibnamefont {Jiang}}, \bibinfo
  {author} {\bibfnamefont {B.}~\bibnamefont {Zhou}}, \bibinfo {author}
  {\bibfnamefont {Z.~J.}\ \bibnamefont {Wang}}, \bibinfo {author}
  {\bibfnamefont {Y.}~\bibnamefont {Zhang}}, \bibinfo {author} {\bibfnamefont
  {H.~M.}\ \bibnamefont {Weng}}, \bibinfo {author} {\bibfnamefont
  {D.}~\bibnamefont {Prabhakaran}}, \bibinfo {author} {\bibfnamefont {S.-K.}\
  \bibnamefont {Mo}}, \bibinfo {author} {\bibfnamefont {H.}~\bibnamefont
  {Peng}}, \bibinfo {author} {\bibfnamefont {P.}~\bibnamefont {Dudin}},
  \bibinfo {author} {\bibfnamefont {T.}~\bibnamefont {Kim}}, \bibinfo {author}
  {\bibfnamefont {M.}~\bibnamefont {Hoesch}}, \bibinfo {author} {\bibfnamefont
  {Z.}~\bibnamefont {Fang}}, \bibinfo {author} {\bibfnamefont {X.}~\bibnamefont
  {Dai}}, \bibinfo {author} {\bibfnamefont {Z.~X.}\ \bibnamefont {Shen}},
  \bibinfo {author} {\bibfnamefont {D.~L.}\ \bibnamefont {Feng}}, \bibinfo
  {author} {\bibfnamefont {Z.}~\bibnamefont {Hussain}},\ and\ \bibinfo {author}
  {\bibfnamefont {Y.~L.}\ \bibnamefont {Chen}},\ }\bibfield  {title} {\bibinfo
  {title} {{A stable three-dimensional topological Dirac semimetal
  Cd$_3$As$_2$}},\ }\href {https://doi.org/10.1038/nmat3990} {\bibfield
  {journal} {\bibinfo  {journal} {Nat. Mater.}\ }\textbf {\bibinfo {volume}
  {13}},\ \bibinfo {pages} {677} (\bibinfo {year} {2014})}\BibitemShut
  {NoStop}%
\bibitem [{\citenamefont {Neupane}\ \emph {et~al.}(2014)\citenamefont
  {Neupane}, \citenamefont {Xu}, \citenamefont {Sankar}, \citenamefont
  {Alidoust}, \citenamefont {Bian}, \citenamefont {Liu}, \citenamefont
  {Belopolski}, \citenamefont {Chang}, \citenamefont {Jeng}, \citenamefont
  {Lin}, \citenamefont {Bansil}, \citenamefont {Chou},\ and\ \citenamefont
  {Hasan}}]{Neupane-Hasan-Cd3As2:2014}%
  \BibitemOpen
  \bibfield  {author} {\bibinfo {author} {\bibfnamefont {M.}~\bibnamefont
  {Neupane}}, \bibinfo {author} {\bibfnamefont {S.-Y.}\ \bibnamefont {Xu}},
  \bibinfo {author} {\bibfnamefont {R.}~\bibnamefont {Sankar}}, \bibinfo
  {author} {\bibfnamefont {N.}~\bibnamefont {Alidoust}}, \bibinfo {author}
  {\bibfnamefont {G.}~\bibnamefont {Bian}}, \bibinfo {author} {\bibfnamefont
  {C.}~\bibnamefont {Liu}}, \bibinfo {author} {\bibfnamefont {I.}~\bibnamefont
  {Belopolski}}, \bibinfo {author} {\bibfnamefont {T.-R.}\ \bibnamefont
  {Chang}}, \bibinfo {author} {\bibfnamefont {H.-T.}\ \bibnamefont {Jeng}},
  \bibinfo {author} {\bibfnamefont {H.}~\bibnamefont {Lin}}, \bibinfo {author}
  {\bibfnamefont {A.}~\bibnamefont {Bansil}}, \bibinfo {author} {\bibfnamefont
  {F.}~\bibnamefont {Chou}},\ and\ \bibinfo {author} {\bibfnamefont {M.~Z.}\
  \bibnamefont {Hasan}},\ }\bibfield  {title} {\bibinfo {title} {{Observation
  of a three-dimensional topological Dirac semimetal phase in high-mobility
  Cd$_3$As$_2$}},\ }\href {https://doi.org/10.1038/ncomms4786} {\bibfield
  {journal} {\bibinfo  {journal} {Nat. Commun.}\ }\textbf {\bibinfo {volume}
  {5}},\ \bibinfo {pages} {3786} (\bibinfo {year} {2014})}\BibitemShut
  {NoStop}%
\bibitem [{\citenamefont {Wang}\ \emph {et~al.}(2019)\citenamefont {Wang},
  \citenamefont {Jo}, \citenamefont {Kuthanazhi}, \citenamefont {Wu},
  \citenamefont {McQueeney}, \citenamefont {Kaminski},\ and\ \citenamefont
  {Canfield}}]{Wang-Canfield:2019}%
  \BibitemOpen
  \bibfield  {author} {\bibinfo {author} {\bibfnamefont {L.-L.}\ \bibnamefont
  {Wang}}, \bibinfo {author} {\bibfnamefont {N.~H.}\ \bibnamefont {Jo}},
  \bibinfo {author} {\bibfnamefont {B.}~\bibnamefont {Kuthanazhi}}, \bibinfo
  {author} {\bibfnamefont {Y.}~\bibnamefont {Wu}}, \bibinfo {author}
  {\bibfnamefont {R.~J.}\ \bibnamefont {McQueeney}}, \bibinfo {author}
  {\bibfnamefont {A.}~\bibnamefont {Kaminski}},\ and\ \bibinfo {author}
  {\bibfnamefont {P.~C.}\ \bibnamefont {Canfield}},\ }\bibfield  {title}
  {\bibinfo {title} {{Single pair of Weyl fermions in the half-metallic
  semimetal EuCd$_2$As$_2$}},\ }\href
  {https://doi.org/10.1103/PhysRevB.99.245147} {\bibfield  {journal} {\bibinfo
  {journal} {Phys. Rev. B}\ }\textbf {\bibinfo {volume} {99}},\ \bibinfo
  {pages} {245147} (\bibinfo {year} {2019})}\BibitemShut {NoStop}%
\bibitem [{\citenamefont {Soh}\ \emph {et~al.}(2019)\citenamefont {Soh},
  \citenamefont {de~Juan}, \citenamefont {Vergniory}, \citenamefont
  {Schr{\"{o}}ter}, \citenamefont {Rahn}, \citenamefont {Yan}, \citenamefont
  {Jiang}, \citenamefont {Bristow}, \citenamefont {Reiss}, \citenamefont
  {Blandy}, \citenamefont {Guo}, \citenamefont {Shi}, \citenamefont {Kim},
  \citenamefont {McCollam}, \citenamefont {Simon}, \citenamefont {Chen},
  \citenamefont {Coldea},\ and\ \citenamefont
  {Boothroyd}}]{Soh-Boothroyd:2019}%
  \BibitemOpen
  \bibfield  {author} {\bibinfo {author} {\bibfnamefont {J.-R.}\ \bibnamefont
  {Soh}}, \bibinfo {author} {\bibfnamefont {F.}~\bibnamefont {de~Juan}},
  \bibinfo {author} {\bibfnamefont {M.~G.}\ \bibnamefont {Vergniory}}, \bibinfo
  {author} {\bibfnamefont {N.~B.~M.}\ \bibnamefont {Schr{\"{o}}ter}}, \bibinfo
  {author} {\bibfnamefont {M.~C.}\ \bibnamefont {Rahn}}, \bibinfo {author}
  {\bibfnamefont {D.~Y.}\ \bibnamefont {Yan}}, \bibinfo {author} {\bibfnamefont
  {J.}~\bibnamefont {Jiang}}, \bibinfo {author} {\bibfnamefont
  {M.}~\bibnamefont {Bristow}}, \bibinfo {author} {\bibfnamefont {P.~A.}\
  \bibnamefont {Reiss}}, \bibinfo {author} {\bibfnamefont {J.~N.}\ \bibnamefont
  {Blandy}}, \bibinfo {author} {\bibfnamefont {Y.~F.}\ \bibnamefont {Guo}},
  \bibinfo {author} {\bibfnamefont {Y.~G.}\ \bibnamefont {Shi}}, \bibinfo
  {author} {\bibfnamefont {T.~K.}\ \bibnamefont {Kim}}, \bibinfo {author}
  {\bibfnamefont {A.}~\bibnamefont {McCollam}}, \bibinfo {author}
  {\bibfnamefont {S.~H.}\ \bibnamefont {Simon}}, \bibinfo {author}
  {\bibfnamefont {Y.}~\bibnamefont {Chen}}, \bibinfo {author} {\bibfnamefont
  {A.~I.}\ \bibnamefont {Coldea}},\ and\ \bibinfo {author} {\bibfnamefont
  {A.~T.}\ \bibnamefont {Boothroyd}},\ }\bibfield  {title} {\bibinfo {title}
  {{Ideal Weyl semimetal induced by magnetic exchange}},\ }\href
  {https://doi.org/10.1103/PhysRevB.100.201102} {\bibfield  {journal} {\bibinfo
   {journal} {Phys. Rev. B}\ }\textbf {\bibinfo {volume} {100}},\ \bibinfo
  {pages} {201102(R)} (\bibinfo {year} {2019})}\BibitemShut {NoStop}%
\bibitem [{\citenamefont {Ma}\ \emph {et~al.}(2019)\citenamefont {Ma},
  \citenamefont {Nie}, \citenamefont {Yi}, \citenamefont {Jandke},
  \citenamefont {Shang}, \citenamefont {Yao}, \citenamefont {Naamneh},
  \citenamefont {Yan}, \citenamefont {Sun}, \citenamefont {Chikina},
  \citenamefont {Strocov}, \citenamefont {Medarde}, \citenamefont {Song},
  \citenamefont {Xiong}, \citenamefont {Xu}, \citenamefont {Wulfhekel},
  \citenamefont {Mesot}, \citenamefont {Reticcioli}, \citenamefont {Franchini},
  \citenamefont {Mudry}, \citenamefont {M{\"{u}}ller}, \citenamefont {Shi},
  \citenamefont {Qian}, \citenamefont {Ding},\ and\ \citenamefont
  {Shi}}]{Ma-Shi:2019}%
  \BibitemOpen
  \bibfield  {author} {\bibinfo {author} {\bibfnamefont {J.-Z.}\ \bibnamefont
  {Ma}}, \bibinfo {author} {\bibfnamefont {S.~M.}\ \bibnamefont {Nie}},
  \bibinfo {author} {\bibfnamefont {C.~J.}\ \bibnamefont {Yi}}, \bibinfo
  {author} {\bibfnamefont {J.}~\bibnamefont {Jandke}}, \bibinfo {author}
  {\bibfnamefont {T.}~\bibnamefont {Shang}}, \bibinfo {author} {\bibfnamefont
  {M.~Y.}\ \bibnamefont {Yao}}, \bibinfo {author} {\bibfnamefont
  {M.}~\bibnamefont {Naamneh}}, \bibinfo {author} {\bibfnamefont {L.~Q.}\
  \bibnamefont {Yan}}, \bibinfo {author} {\bibfnamefont {Y.}~\bibnamefont
  {Sun}}, \bibinfo {author} {\bibfnamefont {A.}~\bibnamefont {Chikina}},
  \bibinfo {author} {\bibfnamefont {V.~N.}\ \bibnamefont {Strocov}}, \bibinfo
  {author} {\bibfnamefont {M.}~\bibnamefont {Medarde}}, \bibinfo {author}
  {\bibfnamefont {M.}~\bibnamefont {Song}}, \bibinfo {author} {\bibfnamefont
  {Y.-M.}\ \bibnamefont {Xiong}}, \bibinfo {author} {\bibfnamefont
  {G.}~\bibnamefont {Xu}}, \bibinfo {author} {\bibfnamefont {W.}~\bibnamefont
  {Wulfhekel}}, \bibinfo {author} {\bibfnamefont {J.}~\bibnamefont {Mesot}},
  \bibinfo {author} {\bibfnamefont {M.}~\bibnamefont {Reticcioli}}, \bibinfo
  {author} {\bibfnamefont {C.}~\bibnamefont {Franchini}}, \bibinfo {author}
  {\bibfnamefont {C.}~\bibnamefont {Mudry}}, \bibinfo {author} {\bibfnamefont
  {M.}~\bibnamefont {M{\"{u}}ller}}, \bibinfo {author} {\bibfnamefont {Y.~G.}\
  \bibnamefont {Shi}}, \bibinfo {author} {\bibfnamefont {T.}~\bibnamefont
  {Qian}}, \bibinfo {author} {\bibfnamefont {H.}~\bibnamefont {Ding}},\ and\
  \bibinfo {author} {\bibfnamefont {M.}~\bibnamefont {Shi}},\ }\bibfield
  {title} {\bibinfo {title} {{Spin fluctuation induced Weyl semimetal state in
  the paramagnetic phase of EuCd$_2$As$_2$}},\ }\href
  {https://doi.org/10.1126/sciadv.aaw4718} {\bibfield  {journal} {\bibinfo
  {journal} {Sci. Adv.}\ }\textbf {\bibinfo {volume} {5}},\ \bibinfo {pages}
  {eaaw4718} (\bibinfo {year} {2019})}\BibitemShut {NoStop}%
\end{thebibliography}%


\begin{thebibliography}{13}%
\makeatletter
\providecommand \@ifxundefined [1]{%
 \@ifx{#1\undefined}
}%
\providecommand \@ifnum [1]{%
 \ifnum #1\expandafter \@firstoftwo
 \else \expandafter \@secondoftwo
 \fi
}%
\providecommand \@ifx [1]{%
 \ifx #1\expandafter \@firstoftwo
 \else \expandafter \@secondoftwo
 \fi
}%
\providecommand \natexlab [1]{#1}%
\providecommand \enquote  [1]{``#1''}%
\providecommand \bibnamefont  [1]{#1}%
\providecommand \bibfnamefont [1]{#1}%
\providecommand \citenamefont [1]{#1}%
\providecommand \href@noop [0]{\@secondoftwo}%
\providecommand \href [0]{\begingroup \@sanitize@url \@href}%
\providecommand \@href[1]{\@@startlink{#1}\@@href}%
\providecommand \@@href[1]{\endgroup#1\@@endlink}%
\providecommand \@sanitize@url [0]{\catcode `\\12\catcode `\$12\catcode
  `\&12\catcode `\#12\catcode `\^12\catcode `\_12\catcode `\%12\relax}%
\providecommand \@@startlink[1]{}%
\providecommand \@@endlink[0]{}%
\providecommand \url  [0]{\begingroup\@sanitize@url \@url }%
\providecommand \@url [1]{\endgroup\@href {#1}{\urlprefix }}%
\providecommand \urlprefix  [0]{URL }%
\providecommand \Eprint [0]{\href }%
\providecommand \doibase [0]{https://doi.org/}%
\providecommand \selectlanguage [0]{\@gobble}%
\providecommand \bibinfo  [0]{\@secondoftwo}%
\providecommand \bibfield  [0]{\@secondoftwo}%
\providecommand \translation [1]{[#1]}%
\providecommand \BibitemOpen [0]{}%
\providecommand \bibitemStop [0]{}%
\providecommand \bibitemNoStop [0]{.\EOS\space}%
\providecommand \EOS [0]{\spacefactor3000\relax}%
\providecommand \BibitemShut  [1]{\csname bibitem#1\endcsname}%
\let\auto@bib@innerbib\@empty
\bibitem [{\citenamefont {Xiao}\ \emph {et~al.}(2010)\citenamefont {Xiao},
  \citenamefont {Chang},\ and\ \citenamefont {Niu}}]{Xiao-Niu:rev-2010}%
  \BibitemOpen
  \bibfield  {author} {\bibinfo {author} {\bibfnamefont {D.}~\bibnamefont
  {Xiao}}, \bibinfo {author} {\bibfnamefont {M.~C.}\ \bibnamefont {Chang}},\
  and\ \bibinfo {author} {\bibfnamefont {Q.}~\bibnamefont {Niu}},\ }\bibfield
  {title} {\bibinfo {title} {{Berry phase effects on electronic properties}},\
  }\href {https://doi.org/10.1103/RevModPhys.82.1959} {\bibfield  {journal}
  {\bibinfo  {journal} {Rev. Mod. Phys.}\ }\textbf {\bibinfo {volume} {82}},\
  \bibinfo {pages} {1959} (\bibinfo {year} {2010})}\BibitemShut {NoStop}%
\bibitem [{\citenamefont {Son}\ and\ \citenamefont
  {Yamamoto}(2013)}]{Son:2013}%
  \BibitemOpen
  \bibfield  {author} {\bibinfo {author} {\bibfnamefont {D.~T.}\ \bibnamefont
  {Son}}\ and\ \bibinfo {author} {\bibfnamefont {N.}~\bibnamefont {Yamamoto}},\
  }\bibfield  {title} {\bibinfo {title} {{Kinetic theory with Berry curvature
  from quantum field theories}},\ }\href
  {https://doi.org/10.1103/PhysRevD.87.085016} {\bibfield  {journal} {\bibinfo
  {journal} {Phys. Rev. D}\ }\textbf {\bibinfo {volume} {87}},\ \bibinfo
  {pages} {085016} (\bibinfo {year} {2013})}\BibitemShut {NoStop}%
\bibitem [{\citenamefont {Stephanov}\ and\ \citenamefont
  {Yin}(2012)}]{Stephanov:2012}%
  \BibitemOpen
  \bibfield  {author} {\bibinfo {author} {\bibfnamefont {M.~A.}\ \bibnamefont
  {Stephanov}}\ and\ \bibinfo {author} {\bibfnamefont {Y.}~\bibnamefont
  {Yin}},\ }\bibfield  {title} {\bibinfo {title} {{Chiral kinetic theory}},\
  }\href {https://doi.org/10.1103/PhysRevLett.109.162001} {\bibfield  {journal}
  {\bibinfo  {journal} {Phys. Rev. Lett.}\ }\textbf {\bibinfo {volume} {109}},\
  \bibinfo {pages} {162001} (\bibinfo {year} {2012})}\BibitemShut {NoStop}%
\bibitem [{\citenamefont {Son}\ and\ \citenamefont
  {Spivak}(2013)}]{Son-Spivak:2013}%
  \BibitemOpen
  \bibfield  {author} {\bibinfo {author} {\bibfnamefont {D.~T.}\ \bibnamefont
  {Son}}\ and\ \bibinfo {author} {\bibfnamefont {B.~Z.}\ \bibnamefont
  {Spivak}},\ }\bibfield  {title} {\bibinfo {title} {{Chiral anomaly and
  classical negative magnetoresistance of Weyl metals}},\ }\href
  {https://doi.org/10.1103/PhysRevB.88.104412} {\bibfield  {journal} {\bibinfo
  {journal} {Phys. Rev. B}\ }\textbf {\bibinfo {volume} {88}},\ \bibinfo
  {pages} {104412} (\bibinfo {year} {2013})}\BibitemShut {NoStop}%
\bibitem [{\citenamefont {Abrikosov}(2017)}]{Abrikosov:book-1988}%
  \BibitemOpen
  \bibfield  {author} {\bibinfo {author} {\bibfnamefont {A.~A.}\ \bibnamefont
  {Abrikosov}},\ }\href {https://books.google.ca/books?id=tTo2DwAAQBAJ} {\emph
  {\bibinfo {title} {{Fundamentals of the Theory of Metals}}}}\ (\bibinfo
  {publisher} {Courier Dover Publications},\ \bibinfo {address} {New York},\
  \bibinfo {year} {2017})\BibitemShut {NoStop}%
\bibitem [{\citenamefont {Lifshitz}\ \emph {et~al.}(1957)\citenamefont
  {Lifshitz}, \citenamefont {Azbel},\ and\ \citenamefont
  {Kaganov}}]{Lifshitz-Kaganov:1957}%
  \BibitemOpen
  \bibfield  {author} {\bibinfo {author} {\bibfnamefont {I.~M.}\ \bibnamefont
  {Lifshitz}}, \bibinfo {author} {\bibfnamefont {M.~I.}\ \bibnamefont
  {Azbel}},\ and\ \bibinfo {author} {\bibfnamefont {M.~I.}\ \bibnamefont
  {Kaganov}},\ }\bibfield  {title} {\bibinfo {title} {{The theory of
  galvanomagnetic effects in metals}},\ }\href
  {http://www.jetp.ac.ru/cgi-bin/dn/e_004_01_0041.pdf} {\bibfield  {journal}
  {\bibinfo  {journal} {JETP}\ }\textbf {\bibinfo {volume} {4}},\ \bibinfo
  {pages} {41} (\bibinfo {year} {1957})}\BibitemShut {NoStop}%
\bibitem [{\citenamefont {Burkov}(2014)}]{Burkov:2014b}%
  \BibitemOpen
  \bibfield  {author} {\bibinfo {author} {\bibfnamefont {A.~A.}\ \bibnamefont
  {Burkov}},\ }\bibfield  {title} {\bibinfo {title} {{Chiral anomaly and
  diffusive magnetotransport in Weyl metals}},\ }\href
  {https://doi.org/10.1103/PhysRevLett.113.247203} {\bibfield  {journal}
  {\bibinfo  {journal} {Phys. Rev. Lett.}\ }\textbf {\bibinfo {volume} {113}},\
  \bibinfo {pages} {247203} (\bibinfo {year} {2014})}\BibitemShut {NoStop}%
\bibitem [{\citenamefont {Parameswaran}\ \emph {et~al.}(2014)\citenamefont
  {Parameswaran}, \citenamefont {Grover}, \citenamefont {Abanin}, \citenamefont
  {Pesin},\ and\ \citenamefont {Vishwanath}}]{Parameswaran-Vishwanath:2014}%
  \BibitemOpen
  \bibfield  {author} {\bibinfo {author} {\bibfnamefont {S.~A.}\ \bibnamefont
  {Parameswaran}}, \bibinfo {author} {\bibfnamefont {T.}~\bibnamefont
  {Grover}}, \bibinfo {author} {\bibfnamefont {D.~A.}\ \bibnamefont {Abanin}},
  \bibinfo {author} {\bibfnamefont {D.~A.}\ \bibnamefont {Pesin}},\ and\
  \bibinfo {author} {\bibfnamefont {A.}~\bibnamefont {Vishwanath}},\ }\bibfield
   {title} {\bibinfo {title} {{Probing the chiral anomaly with nonlocal
  transport in three-dimensional topological semimetals}},\ }\href
  {https://doi.org/10.1103/PhysRevX.4.031035} {\bibfield  {journal} {\bibinfo
  {journal} {Phys. Rev. X}\ }\textbf {\bibinfo {volume} {4}},\ \bibinfo {pages}
  {031035} (\bibinfo {year} {2014})}\BibitemShut {NoStop}%
\bibitem [{\citenamefont {Burkov}(2015)}]{Burkov:rev-2015}%
  \BibitemOpen
  \bibfield  {author} {\bibinfo {author} {\bibfnamefont {A.~A.}\ \bibnamefont
  {Burkov}},\ }\bibfield  {title} {\bibinfo {title} {{Chiral anomaly and
  transport in Weyl metals}},\ }\href
  {https://doi.org/10.1088/0953-8984/27/11/113201} {\bibfield  {journal}
  {\bibinfo  {journal} {J. Phys. Condens. Matter}\ }\textbf {\bibinfo {volume}
  {27}},\ \bibinfo {pages} {113201} (\bibinfo {year} {2015})}\BibitemShut
  {NoStop}%
\bibitem [{\citenamefont {Landau}\ \emph {et~al.}(1984)\citenamefont {Landau},
  \citenamefont {Lifshits},\ and\ \citenamefont {Pitaevskii}}]{Landau:t8}%
  \BibitemOpen
  \bibfield  {author} {\bibinfo {author} {\bibfnamefont {L.~D.}\ \bibnamefont
  {Landau}}, \bibinfo {author} {\bibfnamefont {E.~M.}\ \bibnamefont
  {Lifshits}},\ and\ \bibinfo {author} {\bibfnamefont {L.~P.}\ \bibnamefont
  {Pitaevskii}},\ }\href
  {https://www.elsevier.com/books/electrodynamics-of-continuous-media/landau/978-0-08-057060-0}
  {\emph {\bibinfo {title} {{Electrodynamics of Continuous Media}}}}\ (\bibinfo
   {publisher} {Butterworth-Heinemann},\ \bibinfo {address} {Oxford},\ \bibinfo
  {year} {1984})\BibitemShut {NoStop}%
\bibitem [{\citenamefont {Burkov}(2018)}]{Burkov:2018-OptCond}%
  \BibitemOpen
  \bibfield  {author} {\bibinfo {author} {\bibfnamefont {A.~A.}\ \bibnamefont
  {Burkov}},\ }\bibfield  {title} {\bibinfo {title} {{Dynamical density
  response and optical conductivity in topological metals}},\ }\href
  {https://doi.org/10.1103/PhysRevB.98.165123} {\bibfield  {journal} {\bibinfo
  {journal} {Phys. Rev. B}\ }\textbf {\bibinfo {volume} {98}},\ \bibinfo
  {pages} {165123} (\bibinfo {year} {2018})}\BibitemShut {NoStop}%
\bibitem [{\citenamefont {Arnold}\ \emph {et~al.}(2016)\citenamefont {Arnold},
  \citenamefont {Naumann}, \citenamefont {Wu}, \citenamefont {Sun},
  \citenamefont {Schmidt}, \citenamefont {Borrmann}, \citenamefont {Felser},
  \citenamefont {Yan},\ and\ \citenamefont {Hassinger}}]{Arnold-Felser:2016b}%
  \BibitemOpen
  \bibfield  {author} {\bibinfo {author} {\bibfnamefont {F.}~\bibnamefont
  {Arnold}}, \bibinfo {author} {\bibfnamefont {M.}~\bibnamefont {Naumann}},
  \bibinfo {author} {\bibfnamefont {S.-C.}\ \bibnamefont {Wu}}, \bibinfo
  {author} {\bibfnamefont {Y.}~\bibnamefont {Sun}}, \bibinfo {author}
  {\bibfnamefont {M.}~\bibnamefont {Schmidt}}, \bibinfo {author} {\bibfnamefont
  {H.}~\bibnamefont {Borrmann}}, \bibinfo {author} {\bibfnamefont
  {C.}~\bibnamefont {Felser}}, \bibinfo {author} {\bibfnamefont
  {B.}~\bibnamefont {Yan}},\ and\ \bibinfo {author} {\bibfnamefont
  {E.}~\bibnamefont {Hassinger}},\ }\bibfield  {title} {\bibinfo {title}
  {{Chiral Weyl pockets and Fermi surface topology of the Weyl semimetal
  TaAs}},\ }\href {https://doi.org/10.1103/PhysRevLett.117.146401} {\bibfield
  {journal} {\bibinfo  {journal} {Phys. Rev. Lett.}\ }\textbf {\bibinfo
  {volume} {117}},\ \bibinfo {pages} {146401} (\bibinfo {year}
  {2016})}\BibitemShut {NoStop}%
\bibitem [{\citenamefont {Zhang}\ \emph {et~al.}(2016)\citenamefont {Zhang},
  \citenamefont {Xu}, \citenamefont {Belopolski}, \citenamefont {Yuan},
  \citenamefont {Lin}, \citenamefont {Tong}, \citenamefont {Bian},
  \citenamefont {Alidoust}, \citenamefont {Lee}, \citenamefont {Huang},
  \citenamefont {Chang}, \citenamefont {Chang}, \citenamefont {Hsu},
  \citenamefont {Jeng}, \citenamefont {Neupane}, \citenamefont {Sanchez},
  \citenamefont {Zheng}, \citenamefont {Wang}, \citenamefont {Lin},
  \citenamefont {Zhang}, \citenamefont {Lu}, \citenamefont {Shen},
  \citenamefont {Neupert}, \citenamefont {{Zahid Hasan}},\ and\ \citenamefont
  {Jia}}]{Zhang-Hasan-TaAs:2016}%
  \BibitemOpen
  \bibfield  {author} {\bibinfo {author} {\bibfnamefont {C.-L.}\ \bibnamefont
  {Zhang}}, \bibinfo {author} {\bibfnamefont {S.-Y.}\ \bibnamefont {Xu}},
  \bibinfo {author} {\bibfnamefont {I.}~\bibnamefont {Belopolski}}, \bibinfo
  {author} {\bibfnamefont {Z.}~\bibnamefont {Yuan}}, \bibinfo {author}
  {\bibfnamefont {Z.}~\bibnamefont {Lin}}, \bibinfo {author} {\bibfnamefont
  {B.}~\bibnamefont {Tong}}, \bibinfo {author} {\bibfnamefont {G.}~\bibnamefont
  {Bian}}, \bibinfo {author} {\bibfnamefont {N.}~\bibnamefont {Alidoust}},
  \bibinfo {author} {\bibfnamefont {C.-C.}\ \bibnamefont {Lee}}, \bibinfo
  {author} {\bibfnamefont {S.-M.}\ \bibnamefont {Huang}}, \bibinfo {author}
  {\bibfnamefont {T.-R.}\ \bibnamefont {Chang}}, \bibinfo {author}
  {\bibfnamefont {G.}~\bibnamefont {Chang}}, \bibinfo {author} {\bibfnamefont
  {C.-H.}\ \bibnamefont {Hsu}}, \bibinfo {author} {\bibfnamefont {H.-T.}\
  \bibnamefont {Jeng}}, \bibinfo {author} {\bibfnamefont {M.}~\bibnamefont
  {Neupane}}, \bibinfo {author} {\bibfnamefont {D.~S.}\ \bibnamefont
  {Sanchez}}, \bibinfo {author} {\bibfnamefont {H.}~\bibnamefont {Zheng}},
  \bibinfo {author} {\bibfnamefont {J.}~\bibnamefont {Wang}}, \bibinfo {author}
  {\bibfnamefont {H.}~\bibnamefont {Lin}}, \bibinfo {author} {\bibfnamefont
  {C.}~\bibnamefont {Zhang}}, \bibinfo {author} {\bibfnamefont {H.-Z.}\
  \bibnamefont {Lu}}, \bibinfo {author} {\bibfnamefont {S.-Q.}\ \bibnamefont
  {Shen}}, \bibinfo {author} {\bibfnamefont {T.}~\bibnamefont {Neupert}},
  \bibinfo {author} {\bibfnamefont {M.}~\bibnamefont {{Zahid Hasan}}},\ and\
  \bibinfo {author} {\bibfnamefont {S.}~\bibnamefont {Jia}},\ }\bibfield
  {title} {\bibinfo {title} {{Signatures of the Adler–Bell–Jackiw chiral
  anomaly in a Weyl fermion semimetal}},\ }\href
  {https://doi.org/10.1038/ncomms10735} {\bibfield  {journal} {\bibinfo
  {journal} {Nat. Commun.}\ }\textbf {\bibinfo {volume} {7}},\ \bibinfo {pages}
  {10735} (\bibinfo {year} {2016})}\BibitemShut {NoStop}%
\end{thebibliography}%

\end{document}


\title{Supplemental Material\\
Anomalous electromagnetic field penetration in a Weyl or Dirac semimetal}

\author{P.~O.~Sukhachov}
\email{pavlo.sukhachov@yale.edu}
\affiliation{Department of Physics, Yale University, New Haven, Connecticut 06520, USA}

\author{L.~I.~Glazman}
\affiliation{Department of Physics, Yale University, New Haven, Connecticut 06520, USA}

\maketitle
\tableofcontents

\section{S I. Chiral kinetic theory}
\label{sec:app-CKT}

In this Section, we discuss the derivation of the kinetic equations used in the main text  to describe the quasiparticle dynamics in a Dirac or Weyl semimetal. These equations are necessary to calculate the electric current density $\mathbf{j}(t,\mathbf{r})$, which enters the Maxwell equations
\begin{align}
\label{app-CKT-Maxwell-be}
\bm{\nabla}\times\mathbf{E}(t,\mathbf{r}) &= -\frac{1}{c} \partial_t\mathbf{B}(t,\mathbf{r}), \\
\bm{\nabla}\times\mathbf{B}(t,\mathbf{r}) &= \frac{4\pi}{c}\mathbf{j}(t,\mathbf{r}) +\frac{1}{c}\partial_t\mathbf{E}(t,\mathbf{r})
\label{app-CKT-Maxwell-ee}
\end{align}
describing transverse electromagnetic fields in the semimetal. In addition to Eqs.~(\ref{app-CKT-Maxwell-be}) and (\ref{app-CKT-Maxwell-ee}), one should also take into account the absence of magnetic monopoles $\bm{\nabla}\cdot\mathbf{B}=0$.

In the presence of the static uniform magnetic field $\mathbf{B}_0$ and the Berry curvature $\mathbf{\Omega}_{\alpha}=\mathbf{\Omega}_{\alpha}(\mathbf{p})$, the electric current density $\mathbf{j}(t,\mathbf{r})$ in time-reversal symmetric Weyl semimetals reads as~\cite{Xiao-Niu:rev-2010,Son:2013,Stephanov:2012,Son-Spivak:2013}
\begin{equation}
\label{app-CKT-j-def}
\mathbf{j}(t,\mathbf{r}) = e\sum_{\alpha}^{N_{W}} \int\frac{d^3p}{(2\pi \hbar)^3} \left[\mathbf{v}_{\alpha} -\frac{e}{c}(\mathbf{v}_{\alpha}\cdot\mathbf{\Omega}_{\alpha})\mathbf{B}_0\right] (\partial_{\epsilon_{\alpha}}f_{\alpha}^{(0)}) n_{\alpha}(t,\mathbf{r},\mathbf{p}).
\end{equation}
Here we used the linearized in weak deviations distribution function for electron quasiparticles
\begin{equation}
\label{app-CKT-equation-f-chi}
f_{\alpha}(t,\mathbf{r},\mathbf{p}) =  f_{\alpha}^{(0)}(\mathbf{p})-(\partial_{\epsilon_{\alpha}}f_{\alpha}^{(0)}) n_{\alpha}(t,\mathbf{r},\mathbf{p}),
\end{equation}
where $f_{\alpha}^{(0)}(\mathbf{p})$ is the equilibrium electron distribution function, $\partial_{\epsilon_{\alpha}}$ is the derivative with respect to the quasiparticle energy $\epsilon_{\alpha}$, and $n_{\alpha}(t,\mathbf{r},\mathbf{p}) \sim E(t,\mathbf{r})$ is the perturbed electron distribution, which depends on time $t$, coordinate $\mathbf{r}$, and momentum $\mathbf{p}$. The sum $\sum_{\alpha}^{N_{W}}$ in Eq.~(\ref{app-CKT-j-def}) runs over all $N_{W}$ Weyl nodes, $\mathbf{v}_{\alpha}=\partial_{\mathbf{p}}\epsilon_{\alpha}$ is the quasiparticle group velocity, the effective energy $\epsilon_{\alpha}$ includes the contribution of the orbital magnetic moment~\cite{Xiao-Niu:rev-2010,Son:2013}, {\sl i.e.}, $\epsilon_{\alpha}\to\epsilon_{\alpha}\left[1+ e \left(\mathbf{B}_0\cdot\bm{\Omega}_{\alpha}\right)/c\right]$, $-e$ is the electron charge, and $c$ is the speed of light.

The perturbed distribution function $n_{\alpha}(t,\mathbf{r},\mathbf{p})$ needed to evaluate the electric current in Eq.~(\ref{app-CKT-j-def}) is determined by the linearized equation of the chiral kinetic theory~\cite{Xiao-Niu:rev-2010,Son:2013,Stephanov:2012,Son-Spivak:2013}
\begin{eqnarray}
&&-(\partial_{\epsilon_{\alpha}}f_{\alpha}^{(0)}) \partial_t n_{\alpha}(t,\mathbf{r},\mathbf{p})
+\frac{(\partial_{\epsilon_{\alpha}}f_{\alpha}^{(0)})}{\Theta_{\alpha}(\mathbf{p})} \frac{e}{c}\left[\mathbf{v}_{\alpha}\times \mathbf{B}_0\right]\cdot\partial_{\mathbf{p}} n_{\alpha}(t,\mathbf{r},\mathbf{p})
-\frac{(\partial_{\epsilon}f_{\alpha}^{(0)})}{\Theta_{\alpha}(\mathbf{p})}\left[\mathbf{v}_{\alpha} -\frac{e}{c}(\mathbf{v}_{\alpha}\cdot\mathbf{\Omega}_{\alpha})\mathbf{B}_0\right]\cdot \bm{\nabla} n_{\alpha}(t,\mathbf{r},\mathbf{p})\nonumber\\
&&-I\left[n_{\alpha}\right]=\frac{1}{\Theta_{\alpha}(\mathbf{p})} \left\{e\mathbf{E}_{\alpha}(t,\mathbf{r}) +\frac{e}{c}\left[\mathbf{v}_{\alpha}\times \mathbf{B}_0\right] -\frac{e^2}{c}(\mathbf{E}_{\alpha}(t,\mathbf{r})\cdot\mathbf{B}_0)\mathbf{\Omega}_{\alpha}\right\}\cdot\partial_{\mathbf{p}} f_{\alpha}^{(0)}(\mathbf{p}).
\label{app-CKT-equation-kinetic-equation}
\end{eqnarray}
Here $-e\mathbf{E}_{\alpha}(t,\mathbf{r}) =-e \mathbf{E}(t,\mathbf{r})-\bm{\nabla}\epsilon_{\alpha}$ is the force acting on an electron, $\Theta_{\alpha}(\mathbf{p})=\left[1-e \left(\mathbf{B}_0\cdot \mathbf{\Omega}_{\alpha}\right)/c\right]$ quantifies the renormalization of the phase-space volume~\cite{Xiao-Niu:rev-2010}, and $I\left[n_{\alpha}\right]$ is the collision integral. By using the Fermi golden rule (see, {\sl e.g.}, Ref.~\cite{Abrikosov:book-1988}), the collision integral is defined as
\begin{equation}
\label{app-1-Icoll-inter-1}
I\left[f_{\alpha}(t,\mathbf{r},\mathbf{p})\right] = - \sum_{\beta}^{N_{W}} \int \frac{d^3p^{\prime}}{(2\pi \hbar)^3}\Theta_{\beta}(\mathbf{p}^{\prime}) \frac{2\pi}{\hbar} \left|A_{\alpha, \beta}\right|^2 \delta\left[\epsilon_{\alpha}(p)-\epsilon_{\beta}(p^{\prime})\right] \left[f_{\alpha}(t,\mathbf{r},\mathbf{p})-f_{\beta}(t,\mathbf{r},\mathbf{p}^{\prime})\right],
\end{equation}
where $\left|A_{\alpha, \beta}\right|$ is the scattering amplitude between  Weyl nodes $\alpha$ and $\beta$.

Let us now discuss approximations that will be used to simplify the calculations but still capture qualitatively important effects. First, we can neglect the phase-space volume renormalization $\Theta_{\alpha}(\mathbf{p})$ and the contribution of the magnetic moment to the energy dispersion. These terms lead only to a small contribution to the conductivity tensor compared to the effects of the chiral anomaly~\cite{Son-Spivak:2013}. Second, we ignore the effects of the magnetic field on the intra-valley electron dynamics. For a general orientation of $\mathbf{B}_0$ and $\mathbf{E}(t,z)$, this limits our consideration to classically-weak magnetic fields. Notice that the conductivity tensor component along the direction of a non-quantizing magnetic field is not affected by a cyclotron motion of electrons in materials with a spherical Fermi surface even for classically-strong fields~\cite{Lifshitz-Kaganov:1957}. Finally, we assume that temperature is low compared to the Fermi energy $\mu$.

We apply the above approximation to the collision integral. The equilibrium distribution function depends only on the absolute value of momentum and is given by the standard Fermi-Dirac distribution. By using Eq.~(\ref{app-CKT-equation-f-chi}), we rewrite the collision integral (\ref{app-1-Icoll-inter-1}) as
\begin{eqnarray}
\label{app-1-Icoll-inter-2}
I\left[n_{\alpha}(t,\mathbf{r},\mathbf{p})\right] &\approx& - \sum_{\beta}^{N_{W}} \int \frac{d^3p^{\prime}}{(2\pi \hbar)^3} \frac{2\pi}{\hbar} \left|A_{\alpha, \beta}\right|^2 \delta\left[\epsilon_{\alpha}(p)-\epsilon_{\beta}(p^{\prime})\right] \delta\left[\epsilon_{\alpha}(p)-\mu\right]
\left[n_{\alpha}(t,\mathbf{r},\mathbf{p})- n_{\beta}(t,\mathbf{r},\mathbf{p}^{\prime})\right] \nonumber\\
&=&-\sum_{\beta}^{N_{W}}\frac{n_{\alpha}(t,\mathbf{r},\mathbf{p}) -\overline{n_{\beta}}(t,\mathbf{r})}{\tau_{\alpha,\beta}} \delta\left(\epsilon_{\alpha}-\mu\right).
\end{eqnarray}
Here the perturbed electron distribution in node $\alpha$ averaged over the respective Fermi surface is
\begin{equation}
\label{app-CKT-equation-bar-def}
\overline{n_{\alpha}}(t,\mathbf{r})= \frac{1}{\nu_{\alpha}}\int \frac{d^3p}{(2\pi \hbar)^3} \delta\left(\epsilon_{\alpha}-\mu\right) n_{\alpha}(t,\mathbf{r},\mathbf{p}),
\end{equation}
the relaxation rate is defined as
\begin{equation}
\label{app-1-intra-tau-inter-def}
\frac{1}{\tau_{\alpha, \beta}} = \int \frac{d^3p}{(2\pi \hbar)^3} \frac{2\pi}{\hbar} \left|A_{\alpha, \beta}\right|^2 \delta\left(\epsilon_{\beta}-\mu\right) = \frac{2\pi}{\hbar} \left|A_{\alpha, \beta}\right|^2 \nu_{\beta},
\end{equation}
and the density of states $\nu_{\alpha}$ at the Fermi level is
\begin{equation}
\label{app-CKT-equation-DOS}
\nu_{\alpha} = \int \frac{d^3p}{(2\pi \hbar)^3}\delta\left(\epsilon_{\alpha}-\mu\right) = \frac{\mu^2}{2\pi^2 \hbar^3v_{F,\alpha}^3}.
\end{equation}
In the last expression, we used the linearized dispersion relation for Weyl quasiparticles $\epsilon_{\alpha} =v_{F,\alpha}p$.

By using the above approximations and the collision integral (\ref{app-1-Icoll-inter-2}), Eq.~(\ref{app-CKT-equation-kinetic-equation}) can be rewritten as
\begin{equation}
\left(\partial_t + \mathbf{v}_{\alpha}\cdot\bm{\nabla} \right)n_{\alpha}(t,\mathbf{r},{\mathbf p}_{F,\alpha})
+\sum_{\beta}^{N_{W}}\frac{n_{\alpha}(t,\mathbf{r},\mathbf{p}_{F,\alpha}) -\overline{n_{\beta}}(t,\mathbf{r})}{\tau_{\alpha,\beta}}
= -e\left[\left(\mathbf{v}_{\alpha}\cdot \mathbf{E}(t,\mathbf{r}) \right) -\frac{e}{c}\left(\mathbf{v}_{\alpha}\cdot\mathbf{\Omega}_{\alpha}({\mathbf p}_{F,\alpha})\right) \left(\mathbf{B}_0\cdot \mathbf{E}(t,\mathbf{r}) \right)\right].
\label{app-CKT-equation-kinetic-equation-lin}
\end{equation}
Here $n_{\alpha}(t,\mathbf{r},{\mathbf p}_{F,\alpha})$ is the nonequilibrium part of the distribution function at the Fermi level. Further, $\mathbf{\Omega}_{\alpha}({\mathbf p}_{F,\alpha})$ is the Berry curvature calculated at the Fermi momentum $p_{F,\alpha}=\mu/v_{F,\alpha}$.

To obtain the solutions to Eq.~(\ref{app-CKT-equation-kinetic-equation-lin}), we consider the case of weak internode scattering compared to the intranode one, $\tau_{\alpha,\alpha}\ll\tau_{\alpha,\beta}$, and use $\omega \tau_{\alpha,\alpha}\ll 1$. As one can check by using the numerical parameters given in the main text, this is indeed the case for, {\sl e.g.}, the Weyl semimetal TaAs. Short intranode relaxation time allows us to retain only the first two harmonics in the expansion of the nonequilibrium part of the distribution function:
\begin{equation}
\label{app-CKT-equation-n-expand}
n_{\alpha}(t,\mathbf{r},\mathbf{p}_{F,\alpha}) \approx \overline{n_{\alpha}}(t,\mathbf{r}) + n_{\alpha}^{(1)}(t,\mathbf{r}) \cos{\theta},
\end{equation}
where $\theta$ is the angle between $\mathbf{v}_{\alpha}$ and $\bm{\nabla}$. To obtain $n_{\alpha}^{(1)}(t,\mathbf{r})$, we follow the standard procedure, {\sl i.e.}, by using Eq.~(\ref{app-CKT-equation-kinetic-equation-lin}), we separate the contributions with different powers of $\cos{\theta}$ and solve for $n_{\alpha}^{(1)}(t,\mathbf{r})$. Then, by averaging over the Fermi surface, the kinetic equation
\begin{eqnarray}
\label{app-CKT-equation-n-chi}
\partial_t N_{\alpha}(t,\mathbf{r}) +\left(\bm{\nabla}\cdot\mathbf{j}_{\alpha}(t,\mathbf{r})\right)
= -\sum_{\beta}^{N_{W}}T_{\alpha,\beta}N_{\beta}(t,\mathbf{r}) -e^2\nu_{\alpha} \left(\mathbf{v}_{\Omega,\alpha}\cdot\mathbf{E}(t,\mathbf{r})\right)
\end{eqnarray}
is derived, see, also Refs.~\cite{Burkov:2014b,Parameswaran-Vishwanath:2014,Burkov:rev-2015}. Here
\begin{equation}
\label{app-CKT-N-def}
N_{\alpha}(t,\mathbf{r}) = -e\nu_{\alpha} \overline{n_{\alpha}}(t,\mathbf{r})
\end{equation}
is the valley (or partial) charge density at node $\alpha$ and the valley current density (\ref{app-CKT-j-def}) reads as
\begin{equation}
\label{app-CKT-equation-node-j}
\mathbf{j}_{\alpha}(t,\mathbf{r}) = -\mathbf{v}_{\Omega,\alpha}N_{\alpha}(t,\mathbf{r}) -D_{\alpha}\bm{\nabla}N_{\alpha}(t,\mathbf{r}) +\sigma_{\alpha}\mathbf{E}(t,\mathbf{r}).
\end{equation}
In the above equations, we find it convenient to introduce the anomalous velocity
\begin{equation}
\label{app-vOmega}
\mathbf{v}_{\Omega,\alpha}= \frac{\chi_{\alpha}e\mathbf{B}_0}{4\pi^2 \hbar^2 c \nu_{\alpha}},
\end{equation}
which is determined by the flux of the Berry curvature $\chi_{\alpha}$. Next, the diffusion coefficient is $D_{\alpha}=v_{F,\alpha}^2\tau_{\alpha,\alpha}/3$ and the electric conductivity per valley is $\sigma_{\alpha}=e^2\nu_{\alpha} D_{\alpha}$. Finally, the first term on the right-hand side in Eq.~(\ref{app-CKT-equation-n-chi}) corresponds to the internode scattering with
\begin{equation}
\label{app-impedance-intra-T-def}
T_{\alpha,\beta} = \delta_{\alpha,\beta}\sum_{\gamma}^{N_{W}}\frac{1}{\tau_{\alpha,\gamma}} -\frac{1}{\tau_{\beta,\alpha}},
\end{equation}
where the scattering rate $1/\tau_{\beta,\alpha}$ is defined in Eq.~(\ref{app-1-intra-tau-inter-def}).

\section{S II. Transmission of electromagnetic waves}
\label{sec:app-transmission}

In this Section, we consider the penetration and transmission of electromagnetic waves in a film of a Dirac or Weyl semimetal subject to an external static magnetic field $\mathbf{B}_0$ directed parallel to its surface.

\subsection{S II.A Model setup and key equations}
\label{sec:app-transmission-gen}

Let us start with the model setup and key equations. We use a film geometry where a Weyl or Dirac semimetal has a finite thickness along the $z$-direction, $0\leq z \leq L$. We assume normal incidence of the incoming ($z\leq 0$) electromagnetic wave with electric field
\begin{equation}
\label{app-CKT-E-in-def}
\mathbf{E}_{\rm in}(t,z)=\mathbf{E}_{\rm in} e^{i(kz-\omega t)},
\end{equation}
where $\omega$ is the angular frequency and $k=\omega/c$ is the wave vector of the wave.

In order to find the electric field inside the film, one needs to solve the Maxwell equations (\ref{app-CKT-Maxwell-be}) and (\ref{app-CKT-Maxwell-ee}). As we will show below, the dynamics of the valley charge affects the electric current and, consequently, the electromagnetic field penetration in an external magnetic field. Taking into account that the time-dependence of the reflected $\mathbf{E}_{\rm r}(t,z)$, in-medium $\mathbf{E}(t,z)$, and transmitted $\mathbf{E}_{\rm out}(t,z)$ fields is given by the same function $e^{-i\omega t}$, the Maxwell equations for the transverse fields reduce to
\begin{equation}
\label{app-CKT-sol-Maxwell-general}
\left[\partial_z^2 + 2i q_0^2(\omega) \right] \mathbf{E}(z) = \frac{4\pi i \omega}{c^2}  \sum_{\alpha}^{N_{W}} \mathbf{v}_{\Omega,\alpha} N_{\alpha}(z),
\end{equation}
where $q_0(\omega) =\sqrt{2\pi \sigma_0 \omega}/c$ is the inverse of the skin depth, {\sl i.e.}, $\delta(\omega)=1/q_0(\omega)$~\cite{Landau:t8,Abrikosov:book-1988}. Here we neglected the displacement current for low frequencies $\omega\ll\sigma_0$ with $\sigma_0=\sum_{\alpha}^{N_{W}}\sigma_{\alpha}$ being the static conductivity.

We rewrite the kinetic equation~(\ref{app-CKT-equation-n-chi}) in the following form:
\begin{eqnarray}
\label{app-CKT-equation-n-chi-1}
\sum_{\beta}^{N_{W}}\left[\frac{T_{\alpha,\beta}}{D_{\alpha}} - 2iq_{\alpha}^2(\omega) \delta_{\alpha,\beta} -\delta_{\alpha,\beta}\partial_z^2\right] N_{\beta}(z) = -\frac{e^2\nu_{\alpha}}{D_{\alpha}} \left(\mathbf{v}_{\Omega,\alpha}\cdot\mathbf{E}(z)\right),
\end{eqnarray}
where $q_{\alpha}(\omega)=\sqrt{\omega/(2D_{\alpha})}$.

The in-medium electric field can be represented as a combination of two fields $\mathbf{E}(z)=\mathbf{E}_\parallel(z)+\mathbf{E}_\perp(z)$, parallel and perpendicular to $\mathbf{B}_0$, respectively. The anomaly affects only the former one, $\mathbf{E}_\parallel(z)$, while $\mathbf{E}_\perp(z)$ is independent of $B_0$ for classically-weak fields. Therefore, in what follows, we focus on $\mathbf{E}_\parallel(z)$.

Equations~(\ref{app-CKT-sol-Maxwell-general}) and (\ref{app-CKT-equation-n-chi-1}) should be supplemented with the boundary conditions. Since the tangential components of electric and magnetic fields are continuous at the interfaces, we have the following boundary conditions for electric fields:
\begin{eqnarray}
\label{app-transmission-BC-1}
&&E_{\rm \parallel in}+E_{\rm \parallel r} = E_{\parallel}(z=0),\\
\label{app-transmission-BC-2}
&&ik\left(E_{\rm \parallel in}-E_{\rm \parallel r} \right)= \partial_zE_{\parallel}(z=0),\\
\label{app-transmission-BC-3}
&&E_{\rm \parallel out} e^{ikL} = E_{\parallel}(z=L),\\
\label{app-transmission-BC-4}
&&ikE_{\rm \parallel out} e^{ikL} = \partial_zE_{\parallel}(z=L).
\end{eqnarray}
Here $E_{\rm \parallel r}(t,z)=E_{\rm \parallel r} e^{-i(k z+\omega t)}$ and $E_{\rm \parallel out}(t,z)=E_{\rm \parallel out} e^{i(kz-\omega t)}$ correspond to the reflected and transmitted electric fields at $z\leq0$ and $z\geq L$, respectively. As for the boundary conditions for the valley charge densities, we consider two types of the boundary conditions at the surface of the semimetal:
\begin{equation}
\label{app-CKT-sol-n-alpha-BC}
\mbox{(i)}\quad  N_{\alpha}(z=0,L)=0 \quad \mbox{and} \quad \mbox{(ii)}\quad \partial_z N_{\alpha}(z=0,L)=0,
\end{equation}
namely, Dirichlet and Neumann ones. These conditions correspond, respectively, to the limits of fast and no internode relaxation at the surface. Thus, Eqs.~(\ref{app-CKT-sol-Maxwell-general}) and (\ref{app-CKT-equation-n-chi-1}) together with the boundary conditions  (\ref{app-transmission-BC-1})--(\ref{app-CKT-sol-n-alpha-BC}) form a complete system for the transverse electric field $\mathbf{E}(z)$ and the partial charge densities $N_{\alpha}(z)$.

In the following Sections, we consider two current response regimes: local and nonlocal. While in the case of the nonlocal response the term with the spatial derivative in Eq.~(\ref{app-CKT-equation-n-chi-1}) plays a crucial role, it can be neglected in the local regime. To quantify the strength of the nonlocal effects, we introduce the following quantity:
\begin{equation}
\label{app-eq:xialpha}
\xi_{\alpha}=\frac{q_0(\omega)}{q_{\alpha}(\omega)}= \sqrt{\frac{4\pi}{c^2} \sigma_0 D_{\alpha}}.
\end{equation}
It is defined solely by the material properties and is independent of $\omega$. In the case of sufficiently large frequencies such that the internode scattering can be neglected, nonlocal and local regimes corresponds to $\xi_{\alpha}\gg1$ and $\xi_{\alpha}\ll1$, respectively.

We note that the presence of both diffusion $\propto D_{\alpha}\partial_z^2 $ and internode scattering terms $\propto T_{\alpha,\beta}$ in Eq.~(\ref{app-CKT-equation-n-chi-1}) leads to the coupled set of the diffusion equations. Therefore, to simplify our calculations, we focus on the case of large frequencies compared to the internode scattering rates $T_{\alpha,\beta}$ in the nonlocal regime. The internode scattering can be straightforwardly included in the local regime, see Sec.~S~II.B or if there is a certain symmetry between the Weyl nodes, see Sec.~S~II.D.

\subsection{S II.B Local regime}
\label{sec:app-transmission-local}

In this Section, we consider the transmission and penetration of electromagnetic waves in the local regime $\xi_{\alpha}\ll1$. In this case, the diffusion of the quasiparticles can be neglected, {\sl i.e.}, one can omit the term $\sim \partial_z^2$ in Eq.~(\ref{app-CKT-equation-n-chi-1}) leading to the following kinetic equation:
\begin{eqnarray}
\label{app-transmission-local-n-chi}
\sum_{\beta}^{N_{W}}\left[T_{\alpha,\beta} - i\omega \delta_{\alpha,\beta}\right] N_{\beta}(z) = -e^2\nu_{\alpha} \left(\mathbf{v}_{\Omega,\alpha}\cdot\mathbf{E}(z)\right).
\end{eqnarray}
In order to find a solution to Eq.~(\ref{app-transmission-local-n-chi}), it is convenient to work in the $\tau$-basis in which the scattering rate matrix with the elements $T_{\alpha,\beta}$ is diagonal. This basis is defined as
\begin{equation}
\label{app-transmission-local-T-eigen}
\sum_{\beta}^{N_{W}}T_{\alpha,\beta} \psi_{n,\beta} = \lambda_{n} \psi_{n,\alpha}.
\end{equation}
Here $\psi_{n}$ are eigenfunctions, $\lambda_{n}$ are eigenvalues, and $n=1,\ldots N_{W}$. Multiplying Eq.~(\ref{app-transmission-local-n-chi}) by $\psi_{n,\alpha}^{\dag}$ and summing over $\alpha$, we obtain the following relation:
\begin{equation}
\label{app-transmission-local-n-chi-psi}
\sum_{\alpha}^{N_{W}} \psi_{n,\alpha}^{\dag}N_{\alpha}(z) = -\frac{e^2}{\lambda_{n}-i\omega} \sum_{\alpha}^{N_{W}} \psi_{n,\alpha}^{\dag}\nu_{\alpha} \left(\mathbf{v}_{\Omega,\alpha}\cdot\mathbf{E}(z)\right).
\end{equation}

In the case of the local response, it is possible to find the in-medium and transmitted electric fields in compact form without expanding in weak magnetic fields even for non-symmetric Weyl nodes. By using expression (\ref{app-transmission-local-n-chi-psi}), the right-hand side of Eq.~(\ref{app-CKT-sol-Maxwell-general}) is rewritten as
\begin{equation}
\label{app-transmission-local-E-eq}
\frac{4\pi i \omega}{c^2}  \sum_{n}^{N_{W}}\sum_{\alpha, \beta}^{N_{W}} \mathbf{v}_{\Omega,\alpha} \psi_{n,\alpha} \psi_{n,\beta}^{\dag} N_{\beta}(z)
=-\frac{4\pi i \omega e^2}{c^2} \sum_{n}^{N_{W}}\sum_{\alpha, \beta}^{N_{W}} \nu_{\beta} \mathbf{v}_{\Omega,\alpha} \left(\mathbf{v}_{\Omega,\beta}\cdot\mathbf{E}(z)\right)
\frac{\psi_{n,\alpha}\psi_{n,\beta}^{\dag}}{\lambda_{n}-i\omega}.
\end{equation}
In the case of the field component parallel to the external magnetic field, Eq.~(\ref{app-CKT-sol-Maxwell-general}) reads
\begin{equation}
\label{app-transmission-local-E-par-eq}
\partial_z^2 E_{\parallel}(z) = -\frac{4\pi i \omega }{c^2} \left(\sigma_{0} +e^2 \sum_{n}^{N_{W}}\sum_{\alpha, \beta}^{N_{W}} \nu_{\beta} v_{\Omega,\alpha} v_{\Omega,\beta} \frac{\psi_{n,\alpha}\psi_{n,\beta}^{\dag}}{\lambda_{n}-i\omega} \right)E_{\parallel}(z).
\end{equation}
Therefore, the in-medium electric field in the presence of the chiral anomaly is described by the same equation as at $B_0=0$ but with the following modified conductivity:
\begin{equation}
\label{app-transmission-local-sigma-sol}
\sigma(B_0,\omega) = \sigma_{0}+\sigma_{\rm anom}(B_0,\omega).
\end{equation}
Here the anomalous correction to the conductivity is
\begin{equation}
\label{app-transmission-local-sigma-anom}
\sigma_{\rm anom}(B_0,\omega) = e^2 \sum_{n}^{N_{W}}\sum_{\alpha, \beta}^{N_{W}} \nu_{\beta} v_{\Omega,\alpha} v_{\Omega,\beta} \frac{\psi_{n,\alpha}\psi_{n,\beta}^{\dag}}{\lambda_{n}-i\omega}.
\end{equation}
The result in Eqs.~(\ref{app-transmission-local-sigma-sol}) and (\ref{app-transmission-local-sigma-anom}) generalizes the well-known ``positive" magnetoconductivity (see, {\sl e.g.}, Refs.~\cite{Son-Spivak:2013,Burkov:2018-OptCond}) to the case of nonvanishing frequencies and nonsymmetric Weyl nodes.

The expressions for the penetrated and transmitted fields straightforwardly follow from Eq.~(\ref{app-transmission-local-E-par-eq}) and the boundary conditions (\ref{app-transmission-BC-1})--(\ref{app-transmission-BC-4}). The explicit form of the outgoing electric field at $L\gg c/\mbox{Re}\left\{\sqrt{2\pi \sigma(B_0,\omega) \omega}\right\}$ reads as
\begin{eqnarray}
\label{app-transmission-local-Eslab2-long}
E_{\rm \parallel out}(z=L) &=& (1-i)\sqrt{\frac{2\omega}{\pi \sigma(B_0,\omega)}} e^{-(1-i)L \sqrt{2\pi \sigma(B_0,\omega) \omega}/c} E_{\rm \parallel in}\approx  (1-i)\sqrt{\frac{2\omega}{\pi \sigma_0}} e^{-(1-i)L \sqrt{2\pi \sigma_0\omega}/c} \nonumber\\ &\times&\left[1-\frac{1-i}{2} \frac{\sigma_{\rm anom}(B_0,\omega)}{\sigma_0} \frac{L}{\delta(\omega)} E_{\rm \parallel in}
\right],
\end{eqnarray}
where we expanded in weak $\left|\sigma_{\rm anom}(B_0,\omega)\right|/\sigma_0$ in the last expression.

To compare the result in Eq.~(\ref{app-transmission-local-Eslab2-long}) with its counterparts for the nonlocal regime, see Sec.~S.~II.C, we neglect the internode scattering and introduce the following characteristic magnetic field:
\begin{equation}
\label{app-eq:Balpha}
B_\alpha(\omega)=4\pi\Phi_0\hbar\left(\omega \nu_\alpha\sum_{\beta}^{N_{W}}\nu_{\beta} D_{\beta} \right)^{1/2},
\end{equation}
where $\Phi_0=\pi\hbar c/e$ is the magnetic flux quantum. Then, restoring the real part of the field, we obtain
\begin{equation}
\label{app-transmission-local-Eslab2-long-1}
E_{\rm \parallel out}(t,z=L) \approx 2\sqrt{\frac{\omega}{\pi \sigma_0}} e^{-L/\delta(\omega)} \Bigg[\cos\!\left(\!\frac{L}{\delta(\omega)}-\frac{\pi}{4}-\omega t\!\right)
-\frac{1}{\sqrt{2}} \frac{L}{\delta(\omega)} \sum_{\alpha}^{N_{W}} \frac{B^2_0}{B^2_\alpha(\omega)} \cos\!\left(\!\frac{L}{\delta(\omega)}-\omega t\!\right)
\Bigg] E_{\rm \parallel in}.
\end{equation}

We find it convenient also to separate the amplitude and phase in Eq.~(\ref{app-transmission-local-Eslab2-long}). For this, we expand the argument of exponent in the first expression in Eq.~(\ref{app-transmission-local-Eslab2-long}) and restore the real part of the field. The result reads as
\begin{equation}
\label{app-transmission-local-Eslab2-long-2}
E_{\rm \parallel out}(t,z=L) \approx 2\sqrt{\frac{\omega}{\pi \sigma_0}} \exp{\left\{-\frac{L}{\delta(\omega)}\left[1+\frac{1}{2}\sum_{\alpha}^{N_{W}} \frac{B^2_0}{B^2_\alpha(\omega)}\right]\right\}} \cos\!{\left\{\!\frac{L}{\delta(\omega)} \left[1- \frac{1}{2}\sum_{\alpha}^{N_{W}} \frac{B^2_0}{B^2_\alpha(\omega)}\right]-\frac{\pi}{4}-\omega t\!\right\}} E_{\rm \parallel in}.
\end{equation}
As one can see, the amplitude of the transmitted field always decreases with the magnetic field.

\subsection{S II.C Nonlocal regime}
\label{sec:app-transmission-iter}

Let us discuss the transmission of electromagnetic waves in the strongly nonlocal regime $\xi_{\alpha}\gg1$. In the case of non-quantizing magnetic fields, it is reasonable to solve Eqs.~(\ref{app-CKT-sol-Maxwell-general}) and (\ref{app-CKT-equation-n-chi-1}) iteratively in the magnetic field, {\sl i.e.}, in $\mathbf{v}_{\Omega,\alpha} \propto \mathbf{B}_0$. We start with the case $B_0=0$. Solving Eq.~(\ref{app-CKT-sol-Maxwell-general}) with the vanishing right-hand side and using the boundary conditions (\ref{app-transmission-BC-1})--(\ref{app-transmission-BC-4}), we obtain the following expressions for the in-medium, reflected, and transmitted fields:
\begin{eqnarray}
\label{app-transmission-iter-Ein-0-1}
E_{\parallel}^{(0)}(z) &=& 2kE_{\rm \parallel in} \frac{k \sin{\left[(1+i)(L-z)q_0(\omega)\right]} -(1-i) q_0(\omega) \cos{\left[(1+i)(L-z) q_0(\omega)\right]}}{\left[k^2+2iq_0^2(\omega)\right]\sin{\left[(1+i)Lq_0(\omega)\right]} -2(1-i)kq_0(\omega) \cos{\left[(1+i)L q_0(\omega)\right]}},\\
\label{app-transmission-iter-Er-0}
E_{\rm \parallel r}^{(0)} &=& E_{\rm \parallel in}\frac{\left[k^2-2iq_0^2(\omega)\right] \sin{\left[(1+i)L q_0(\omega)\right]}}{\left[k^2+2iq_0^2(\omega)\right]\sin{\left[(1+i)Lq_0(\omega)\right]} -2(1-i)kq_0(\omega) \cos{\left[(1+i)L q_0(\omega)\right]}},\\
\label{app-transmission-iter-Etran-0}
E_{\rm \parallel out }^{(0)} &=& -E_{\rm \parallel in} \frac{2(1-i)k q_0(\omega)}{(k^2+2iq_0^2(\omega))\sin{\left[(1+i)Lq_0(\omega)\right]} -2(1-i)kq_0(\omega) \cos{\left[(1+i)Lq_0(\omega)\right]} }.
\end{eqnarray}

The next step in determining the effects of the chiral anomaly in the penetration and transmission of electromagnetic waves is to substitute $E_{\parallel}^{(0)}(z)$ into the right-hand side of Eq.~(\ref{app-CKT-equation-n-chi-1}) and find $N_{\alpha}^{(1)}\propto v_{\Omega,\alpha}$. In the case of large frequencies compared to the internode scattering rates, the latter satisfies the following equation:
\begin{equation}
\label{app-transmission-iter-n-chi}
\left[\partial_z^2 +2iq_{\alpha}^2(\omega)\right] N_{\alpha}^{(1)}(z) = \frac{e^2 \nu_{\alpha} v_{\Omega,\alpha}}{D_{\alpha}}E_{\parallel}^{(0)}(z)
\end{equation}
with the boundary condition (\ref{app-CKT-sol-n-alpha-BC}).
The obtained valley charge density is then substituted into Eq.~(\ref{app-CKT-sol-Maxwell-general}), {\sl i.e.},
\begin{equation}
\label{app-transmission-iter-Maxwell}
\left[\partial_z^2 + 2i q_0^2(\omega) \right] E_{\parallel}^{(2)}(z) = \frac{4\pi i \omega}{c^2}  \sum_{\alpha}^{N_{W}} v_{\Omega,\alpha} N_{\alpha}^{(1)}(z)
\end{equation}
and the resulting anomalous correction to the electric field $E_{\parallel}^{(2)}(z)$ is found.

While being straightforward, the outlined procedure leads to cumbersome results in a general case. Therefore, in what follows, we employ a few approximations. We use $k\ll q_0(\omega)$, which agrees with the assumption $\omega\ll\sigma_0$, and consider films of thickness that exceeds the normal-skin depth, {\sl i.e.}, $L\gg1/q_0(\omega)$. (The case of thin films $L\ll1/q_0(\omega)$ is addressed at the end of this Section.) In this case, the electric field is concentrated in the skin layer $0\leq z \lesssim 1/q_0(\omega)$, {\sl i.e.},
\begin{equation}
\label{app-transmission-iter-E0}
E_{\parallel}^{(0)}(z) = (1-i)\frac{k}{q_0(\omega)} e^{-z q_0(\omega)} e^{iz q_0(\omega)}E_{\parallel\rm in}
\end{equation}
at $B_0=0$. Near the boundary $z=L$, we derive the following outgoing field:
\begin{equation}
\label{app-transmission-iter-E0-out}
E_{\rm \parallel out}^{(0)}(z=L) = 2(1-i)\frac{k}{q_0(\omega)} e^{-L q_0(\omega)} e^{iL q_0(\omega)}E_{\parallel\rm in}.
\end{equation}
These expressions can be straightforwardly obtained by expanding Eqs.~(\ref{app-transmission-iter-Ein-0-1}) and (\ref{app-transmission-iter-Etran-0}).

To find the spatial distribution of the valley charge density, we use the fact that the right-hand side of Eq.~(\ref{app-transmission-iter-n-chi}) is non-negligible only in the skin layer, {\sl i.e.}, for $0\leq z \leq z_0$. Here we define $z_0\sim 1/q_0(\omega)$ as a value of the $z$-coordinate for which $e^{-z_0q_0(\omega)}\ll1$. We first find the charge density in the skin layer and then use the corresponding result as the boundary condition for the charge density at $z_0 \lesssim z \leq L$. Retaining only the spatial derivative and integrating Eq.~(\ref{app-transmission-iter-n-chi}) over $z$ from $z$ to $z_0$, we obtain
\begin{equation}
\label{app-transmission-iter-n-1-int}
\partial_z N_{\alpha}^{(1)}(z=z_0) -\partial_z N_{\alpha}^{(1)}(z) = (1+i)\frac{e^2 \nu_{\alpha} v_{\Omega,\alpha}}{2D_{\alpha} q_0(\omega)} E_{\parallel}^{(0)}(z=0).
\end{equation}
In the case of the Dirichlet boundary conditions, we integrate Eq.~(\ref{app-transmission-iter-n-1-int}) over $z$ from $0$ to $z_0$. The result reads as
\begin{equation}
\label{app-transmission-iter-n-1-BC-i}
\mbox{(i)}\quad z_0\partial_z N_{\alpha}^{(1)}(z_0) - N_{\alpha}^{(1)}(z_0) = i\frac{e^2 \nu_{\alpha} v_{\Omega,\alpha}}{2D_{\alpha} q_0^2(\omega)} E_{\parallel}^{(0)}(z=0).
\end{equation}
Here the first term on the left-hand side is negligible at $\xi_{\alpha}\gg1$.

In the case of the Neumann boundary conditions, by setting $z=0$ in Eq.~(\ref{app-transmission-iter-n-1-int}), we obtain the following boundary condition at $z=z_0$:
\begin{equation}
\label{app-transmission-iter-n-1-BC-ii}
\mbox{(ii)}\quad \partial_z N_{\alpha}^{(1)}(z_0)  = (1+i)\frac{e^2 \nu_{\alpha} v_{\Omega,\alpha}}{2D_{\alpha} q_0(\omega)} E_{\parallel}^{(0)}(z=0).
\end{equation}

Next, we solve the kinetic equation (\ref{app-transmission-iter-n-chi}) with the vanishing right-hand side for $z_0\leq z \leq L$. We use the boundary conditions (\ref{app-CKT-sol-n-alpha-BC}) at $z=L$,
as well as the boundary condition (\ref{app-transmission-iter-n-1-BC-i}) or (\ref{app-transmission-iter-n-1-BC-ii}) at $z=z_0\sim 1/q_0(\omega)$. The corresponding solutions for $L-z \gg 1/q_0(\omega)$ read as
\begin{eqnarray}
\label{app-transmission-iter-n-2-i}
\mbox{(i)}\quad N_{\alpha}^{(1)}(z) &=& -i\frac{e^2 \nu_{\alpha} v_{\Omega,\alpha}}{2q_0^2(\omega) D_{\alpha}} \frac{\sin{\left[(1+i)(L-z)q_{\alpha}(\omega)\right]}}{\sin{\left[(1+i)Lq_{\alpha}(\omega)\right]}} E_\parallel^{(0)}(0) ,\\
\label{app-transmission-iter-n-2-ii}
\mbox{(ii)}\quad N_{\alpha}^{(1)}(z) &=& \frac{e^2 \nu_{\alpha} v_{\Omega,\alpha}}{2q_{\alpha}(\omega) q_0(\omega) D_{\alpha}} \frac{\cos{\left[(1+i)(L-z)q_{\alpha}(\omega)\right]}}{\sin{\left[(1+i)Lq_{\alpha}(\omega)\right]}} E_\parallel^{(0)}(0)
\end{eqnarray}
for the Dirichlet and Neumann boundary conditions and are given in the main text.

Having determined the partial charge density $N_{\alpha}^{(1)}(z)$, let us find the anomalous corrections to the electric fields $E_{\parallel}^{(2)}(z)$ and $E_{\rm \parallel out}^{(2)}(z)$. Here $E_{\parallel}^{(2)}(z)$ follows from Eq.~(\ref{app-transmission-iter-Maxwell}) and $E_{\rm \parallel out}^{(2)}(z)$ is found from Eqs.~(\ref{app-transmission-BC-3}) and (\ref{app-transmission-BC-4}). The general form of the solution to Eq.~(\ref{app-transmission-iter-Maxwell}) reads as
\begin{equation}
\label{app-transmission-iter-E2}
E_{\parallel}^{(2)}(z) = \frac{2\pi \omega}{c^2q_0^2(\omega)} \sum_{\alpha}^{N_{W}} v_{\Omega,\alpha} N_{\alpha}^{(1)}(z) + C_1 e^{-zq_0(\omega)} e^{izq_0(\omega)}+C_2 e^{zq_0(\omega)} e^{-izq_0(\omega)}
\end{equation}
for $\xi_{\alpha}\gg1$. Notice that the term $C_1 e^{-zq_0(\omega)} e^{izq_0(\omega)}$ can be neglected compared to the other terms at $z\gg1/q_{0}(\omega)$ and $\xi_{\alpha}\gg1$. Using the boundary conditions (\ref{app-transmission-BC-3}) and (\ref{app-transmission-BC-4}), we obtain
\begin{eqnarray}
\label{app-transmission-iter-C2}
C_2 &=& \frac{1+i}{2q_0(\omega)}e^{-Lq_0(\omega)}e^{iLq_0(\omega)} \frac{2\pi \omega}{c^2q_0^2(\omega)} \sum_{\alpha}^{N_{W}} v_{\Omega,\alpha} \partial_zN_{\alpha}^{(1)}(z=L),\\
\label{app-transmission-iter-Eslab2}
E_{\parallel}^{(2)}(z) &=& \frac{2\pi \omega}{c^2 q_0^2(\omega)} \sum_{\alpha}^{N_{W}} v_{\Omega,\alpha} \left[N_{\alpha}^{(1)}(z) -\frac{1+i}{2q_0(\omega)} e^{-(L-z)q_0(\omega)} e^{i(L-z)q_0(\omega)} \partial_zN_{\alpha}^{(1)}(z=L)\right].
\end{eqnarray}
The transmitted field follows from Eq.~(\ref{app-transmission-BC-3}), {\sl i.e.}, $E_{\rm \parallel out}^{(2)}(z=L)=E_{\parallel}^{(2)}(z=L)$.

As in the main text, we consider two cases: a film thick, $L\gg 1/q_{\alpha}(\omega)$, or thin, $L\ll 1/q_{\alpha}(\omega)$, compared to the diffusion lengths. Let us start with the former case. By using the partial charge densities (\ref{app-transmission-iter-n-2-i}) and (\ref{app-transmission-iter-n-2-ii}), we present the explicit expressions for the anomalous part of the transmitted electric field $E_{\rm out}^{(2)}(z=L)$ for $L\gg 1/q_{\alpha}(\omega)$:
\begin{eqnarray}
\label{app-transmission-iter-thick-Eout2-i}
\mbox{(i)}\,\,  E_{\rm \parallel out}^{(2)}(z=L) &=& -\frac{2\pi \omega}{c^2q_0^2(\omega)} \sum_{\alpha}^{N_{W}}v_{\Omega,\alpha} \frac{(1+i)}{2q_0(\omega)} \partial_zN_{\alpha}^{(1)}(z=L)
= -(1+i) \frac{4\pi e^2}{c^2} \frac{k}{q_0^3(\omega)} \sum_{\alpha}^{N_{W}} \nu_{\alpha} v_{\Omega,\alpha}^2 \frac{1}{\xi_{\alpha}^{3}} e^{-Lq_{\alpha}(\omega)} e^{iLq_{\alpha}(\omega)} E_{\rm \parallel in},\nonumber\\
\\
\label{app-transmission-iter-thick-Eout2-ii}
\mbox{(ii)}\,\,  E_{\rm \parallel out}^{(2)}(z=L) &=& \frac{2\pi \omega}{c^2q_0^2(\omega)} \sum_{\alpha}^{N_{W}} v_{\Omega,\alpha} N_{\alpha}^{(1)}(z=L)
= -(1+i) \frac{4\pi e^2}{c^2} \frac{k}{q_0^3(\omega)}
\sum_{\alpha}^{N_{W}} \nu_{\alpha} v_{\Omega,\alpha}^2 \frac{1}{\xi_{\alpha}}  e^{-Lq_{\alpha}(\omega)} e^{iLq_{\alpha}(\omega)} E_{\rm \parallel in}.
\end{eqnarray}
While the results in Eqs.~(\ref{app-transmission-iter-thick-Eout2-i}) and (\ref{app-transmission-iter-thick-Eout2-ii}) have the same form, there is an additional small prefactor $1/\xi_{\alpha}^2$ in the case of the Dirichlet boundary conditions that originates from the suppression of $N_{\alpha}(z)$ near the boundaries. Such suppression is absent for the Neumann boundary conditions where a uniform valley charge density is allowed.
Furthermore, compared to the nonanomalous part of the transmitted field (\ref{app-transmission-iter-E0-out}), the decay rates of the anomalous parts are much smaller in the nonlocal regime $\xi_{\alpha}\gg1$ or, equivalently, at $q_0(\omega)\gg q_{\alpha}(\omega)$.

In the case of thin compared to the diffusion lengths films, {\sl i.e.}, at $L\ll 1/q_{\alpha}(\omega)$, we have the following anomalous corrections to the transmitted electric fields:
\begin{eqnarray}
\label{app-transmission-iter-thin-Eout2-i}
\mbox{(i)}\quad E_{\rm \parallel out}^{(2)}(z=L) &=& -\frac{2\pi \omega}{c^2q_0^2(\omega)} \sum_{\alpha}^{N_{W}}v_{\Omega,\alpha} \frac{(1+i)}{2q_0(\omega)} \partial_zN_{\alpha}^{(1)}(z=L)
=-i\frac{2\pi e^2}{c^2} \frac{k}{Lq_0^4(\omega)} \sum_{\alpha}^{N_{W}} \nu_{\alpha} v_{\Omega,\alpha}^2 \frac{1}{\xi_{\alpha}^2} E_{\rm \parallel in},\\
\label{app-transmission-iter-thin-Eout2-ii}
\mbox{(ii)}\quad E_{\rm \parallel out}^{(2)}(z=L) &=& \frac{2\pi \omega}{c^2q_0^2(\omega)} \sum_{\alpha}^{N_{W}} v_{\Omega,\alpha} N_{\alpha}^{(1)}(z=L)
=-i\frac{2\pi e^2}{c^2} \frac{k}{Lq_0^4(\omega)} \sum_{\alpha}^{N_{W}} \nu_{\alpha} v_{\Omega,\alpha}^2  E_{\rm \parallel in}.
\end{eqnarray}
It is important to note that instead of the exponential scaling with the thickness as in the case $L\gg 1/q_{\alpha}(\omega)$, we have a $1/L$ decay for $L\ll 1/q_{\alpha}(\omega)$. This scaling occurs because the valley charge, which is created in the skin layer, spreads over the entire film due to diffusion and, therefore, acquires a $\propto1/L$ density.

Let us compare the transmitted electric field in the local and nonlocal response regimes. Unlike the case of the local response discussed in Sec.~S~II.B, the frequency dependence of the anomalous corrections in the nonlocal regime is qualitatively different, {\sl cf.} Eqs.~(\ref{app-transmission-local-Eslab2-long-1}) and (\ref{app-transmission-iter-thick-Eout2-i})-- (\ref{app-transmission-iter-thin-Eout2-ii}). Furthermore, the amplitude of the transmitted field in the nonlocal regime might be enhanced compared to the case $B_0=0$. On the other hand, the transmitted electric field is always reduced by the chiral anomaly in the local regime, see Eq.~(\ref{app-transmission-local-Eslab2-long-2}). As we discuss in the main text, this might allow one to identify the nonlocality induced by the chiral anomaly.

Finally, let us address the case of thin films with $L\ll 1/q_0(\omega)$ and $L\ll 1/q_{\alpha}(\omega)$. In the absence of the magnetic field, the in-medium electric field can be assumed spatially uniform and equivalent to $E_{\rm \parallel in}$. Indeed, expanding the result in Eq.~(\ref{app-transmission-iter-Ein-0-1}) up to the first order in $L q_0(\omega)$ and using $k\ll q_0(\omega)$, we obtain
\begin{equation}
\label{app-transmission-thin2-iter-Ein-0}
E_{\parallel}^{(0)}(z) \approx \left[1 + i\frac{Lq_0^2(\omega)}{2k}\right] E_{\rm \parallel in},
\end{equation}
where $Lq_0^2(\omega)/k\ll1$.

The corresponding valley charge density $N_{\alpha}^{(1)}$ can be straightforwardly found from Eq.~(\ref{app-transmission-iter-n-chi}) with the boundary conditions (\ref{app-CKT-sol-n-alpha-BC}). In the case under consideration, we have
\begin{eqnarray}
\label{app-transmission-thin2-iter-N-i}
\mbox{(i)}\quad N_{\alpha}^{(1)}(z) &=& -\frac{e^2 \nu_{\alpha} v_{\Omega,\alpha}z(L-z)}{2 D_{\alpha}} E_{\parallel}^{(0)}(z),\\
\label{app-transmission-thin2-iter-N-ii}
\mbox{(ii)}\quad N_{\alpha}^{(1)}(z) &=& -i\frac{e^2 \nu_{\alpha} v_{\Omega,\alpha}}{2q_{\alpha}^2(\omega) D_{\alpha}} E_{\parallel}^{(0)}(z).
\end{eqnarray}
While the result for the Neumann boundary conditions (\ref{app-transmission-thin2-iter-N-ii}) can be derived by ignoring the spatial derivative in Eq.~(\ref{app-transmission-iter-n-chi}), the nontrivial valley charge density for $N_{\alpha}^{(1)}(z=0,L)$ vanishes in this approximation. Therefore, in order to obtain the expression in Eq.~(\ref{app-transmission-thin2-iter-N-i}), we solved  Eq.~(\ref{app-transmission-iter-n-chi}) with the boundary conditions (\ref{app-CKT-sol-n-alpha-BC}) and expanded the obtained result in small $L q_{\alpha}(\omega)$. Notice that the valley charge density at $N_{\alpha}^{(1)}(z=0,L)$ is much smaller, $\sim \left[Lq_{\alpha}(\omega)\right]^2$, than for $\partial_zN_{\alpha}^{(1)}(z=0,L)$, {\sl cf.} Eqs.~(\ref{app-transmission-thin2-iter-N-i}) and (\ref{app-transmission-thin2-iter-N-ii}).

The anomalous corrections to the in-medium $E_{\parallel}^{(2)}(z)$ and transmitted $E_{\rm \parallel out}^{(2)}(z)$ electric fields are obtained by substituting the valley charge density (\ref{app-transmission-thin2-iter-N-i}) or (\ref{app-transmission-thin2-iter-N-ii}) into the right-hand side of Eq.~(\ref{app-transmission-iter-Maxwell}), solving the obtained equation, and using the boundary conditions (\ref{app-transmission-BC-1})--(\ref{app-transmission-BC-4}). The final result for the transmitted electric field reads as
\begin{eqnarray}
\label{app-transmission-thin2-iter-E2-i}
\mbox{(i)}\quad E_{\rm \parallel out}^{(2)}(z=L) &=& \left\{1 - \frac{Lq_0^2(\omega)}{k} -\frac{\pi e^2 L}{3c^2k} \left[Lq_0(\omega)\right]^2 \sum_{\alpha}^{N_{W}}\nu_{\alpha} v_{\Omega,\alpha}^2 \frac{1}{\xi_{\alpha}^2}\right\} E_{\rm \parallel in},\\
\label{app-transmission-thin2-iter-E2-ii}
\mbox{(ii)}\quad E_{\rm \parallel out}^{(2)}(z=L) &=& \left[1 -\frac{Lq_0^2(\omega)}{k} -i \frac{4\pi e^2 L}{c^2k}\sum_{\alpha}^{N_{W}}\nu_{\alpha} v_{\Omega,\alpha}^2\right] E_{\rm \parallel in}.
\end{eqnarray}
It is clear that the anomalous correction to the transmitted field for the Dirichlet boundary conditions (\ref{app-transmission-thin2-iter-E2-i}) is strongly suppressed compared to its counterpart for $\partial_zN_{\alpha}^{(1)}(z=0,L)$. This is indeed expected because the role of the boundaries is very strong in thin films.

\subsection{S II.D Internode scattering rate in a model with symmetric Weyl nodes}
\label{sec:app-transmission-tau5}

Let us discuss the role of the internode scattering. We use a simplified model where there is a symmetry between the pairs ($\alpha,-\alpha$) of the Weyl nodes with opposite topological charges, $\chi_{-\alpha}=-\chi_{\alpha}$, and the nodes are well separated. This allows us to include only the intranode scattering and the scattering between the Weyl nodes within the same pair ($\alpha,-\alpha$). In this case, the scattering rate matrix becomes block diagonal. Its nontrivial eigenvalues are $1/\tau_{5,\alpha}\equiv 2/\tau_{\alpha,-\alpha}$. Indeed, it is straightforward to see from Eq.~(\ref{app-impedance-intra-T-def}) that, under assumptions at hand, the only nonzero components of the scattering rate matrix are $T_{\alpha,\alpha}=-T_{\alpha,-\alpha}=-1/\tau_{\alpha,-\alpha}$.

In the case of the nonlocal response and symmetric Weyl nodes, we find it convenient to introduce the valley-even $N_{\alpha}^{\rm (even)}(z)$ and valley-odd (imbalance) $N_{\alpha}^{\rm (odd)}(z)$ charge densities for the pair of the nodes ($\alpha,-\alpha$). Adding and subtracting the kinetic equation (\ref{app-CKT-equation-n-chi-1}) for the nodes ($\alpha,-\alpha$), we derive
\begin{eqnarray}
\label{app-transmission-tau5-even}
&&\left[2iq_{\alpha}^{2}(\omega)  +\partial_z^2 \right] N_{\alpha}^{\rm (even)}(z) =0,\\
\label{app-transmission-tau5-odd}
&&\left[\frac{1}{\tau_{5,\alpha}} -2i D_{\alpha}q_{\alpha}^{2}(\omega) -D_{\alpha}\partial_z^2\right] N_{\alpha}^{\rm (odd)}(z) = -2e^2 \nu_{\alpha} \left(\mathbf{v}_{\Omega,\alpha}\cdot\mathbf{E}(z)\right).
\end{eqnarray}
It is clear that Eq.~(\ref{app-transmission-tau5-even}) has only trivial solutions for the boundary conditions (\ref{app-CKT-sol-n-alpha-BC}). On the other hand, Eq.~(\ref{app-transmission-tau5-odd}) has the same structure as Eq.~(\ref{app-transmission-iter-n-chi}) where the internode scattering was neglected. Therefore, one can use the results obtained in Sec.~S~II.C
with the following replacement:
\begin{equation}
\label{app-transmission-tau5-repl}
q_{\alpha}^2(\omega) \to q_{\alpha}^2(\omega) -\frac{1}{2iD_{\alpha}\tau_{5,\alpha}}.
\end{equation}
Then, the transmitted electric field follows from Eqs.~(\ref{app-transmission-iter-thick-Eout2-i}) and (\ref{app-transmission-iter-thick-Eout2-ii}) or Eqs.~(\ref{app-transmission-iter-thin-Eout2-i}) and (\ref{app-transmission-iter-thin-Eout2-ii}) by multiplying the summands in the anomalous parts by $\omega \tau_{5,\alpha}/\left(i+\omega\tau_{5,\alpha} \right)$ as well as replacing $\xi_{\alpha}\to \tilde{\xi}_{\alpha}=\xi_{\alpha} \sqrt{\omega\tau_{5,\alpha}/\left(i+\omega \tau_{5,\alpha}\right)}$. Notice that, after performing these replacements, the effective decay length in Eqs.~(\ref{app-transmission-iter-thick-Eout2-i}) and (\ref{app-transmission-iter-thick-Eout2-ii}) is determined by $\sqrt{2D_{\alpha}}/\mbox{Re}\left[(1-i)\sqrt{\omega+i/\tau_{5,\alpha}}\right]$ instead of $\xi_{\alpha}/q_0(\omega)$. This limits the region of applicability of our approximations, {\sl i.e.}, decreases the parameter range where the nonlocal current response develops under the conditions of the normal skin effect.

As for the local response considered in Sec.~S~II.B, the anomalous part of the conductivity in Eq.~(\ref{app-transmission-local-sigma-anom}) can be simplified as
\begin{equation}
\label{app-transmission-local-sigma-anom-symmetric}
\sigma_{\rm anom}(\omega)=e^2 \sum_{\alpha}^{N_{W}} \frac{\tau_{5,\alpha}\nu_{\alpha} v_{\Omega,\alpha}^2}{1-i\omega\tau_{5,\alpha}}.
\end{equation}
This expression follows from the fact that for each pair of the nodes ($\alpha,-\alpha$), there is one trivial and one nontrivial eigenvalue, {\sl i.e.}, $0$ and $1/\tau_{5,\alpha}$. It is straightforward to show that trivial eigenvalues do not contribute to the anomalous part of the conductivity. The result in Eq.~(\ref{app-transmission-local-sigma-anom-symmetric}) agrees with the positive magnetoconductivity derived at $\omega=0$ and $\tau_{5,\alpha}=\tau_{5}$ in, {\sl e.g.}, Ref.~\cite{Son-Spivak:2013}.

In order to estimate the effect of the internode scattering, it is sufficient to use the linearized low-energy Hamiltonian in the vicinity of the Weyl node $\alpha$: $H_{\alpha} =\chi_{\alpha} v_{F} \left(\mathbf{p}\cdot\bm{\sigma}\right)$, where $\bm{\sigma}$ is the vector of the Pauli matrices acting in the pseudospin space and $\chi_{\pm\alpha}=\pm1$. Because the Weyl nodes are symmetric, nodal indices will be omitted henceforth.

\begin{figure}[b]
\centering
\includegraphics[width=0.45\textwidth]{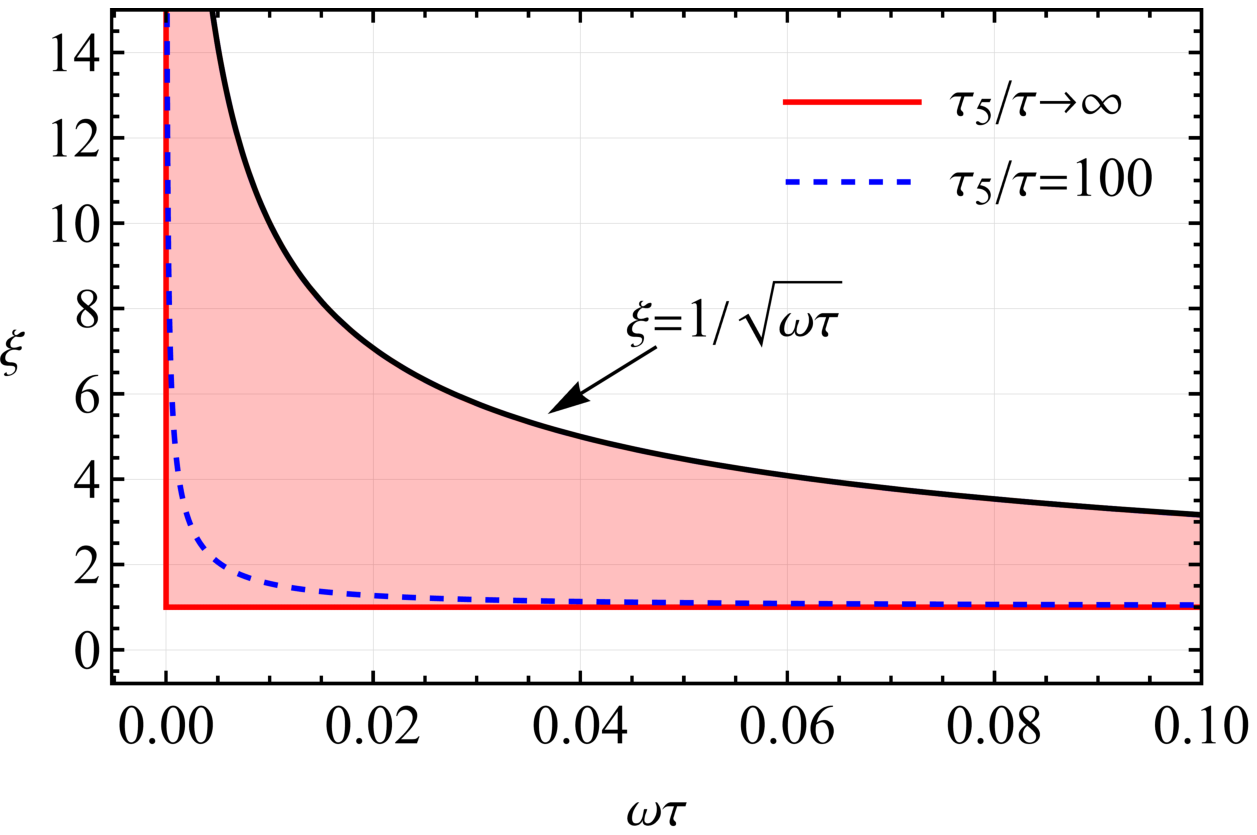}
\caption{
The parameter range where the nonlocal regime for the conditions of the normal skin effect can be realized, {\sl i.e.}, where the conditions $\xi \ll1/\sqrt{\omega\tau}$ and $\xi \gg\mbox{Re}\left[(1-i)\sqrt{1+i/(\omega\tau_{5})}\right]$ hold. The red shaded region denotes the largest parameter range realized at $\tau_{5}/\tau\to\infty$. The black solid line denotes the upper boundary of the region defined by $\xi =1/\sqrt{\omega\tau}$, which is equivalent to $\delta(\omega)=\ell$ with $\ell=v_F\tau$ being the mean free path. The blue dashed line corresponds to the lower boundary of the region at $\tau_{5}/\tau=100$.
}
\label{fig:app-Numerics-parameters}
\end{figure}

We present the parameter range where the nonlocal current response regime for the conditions of the normal skin effect can be realized in Fig.~\ref{fig:app-Numerics-parameters}. In particular, we require that $\ell \ll \delta(\omega)$ and $\delta(\omega)\ll \sqrt{2D}/\mbox{Re}\left[(1-i)\sqrt{\omega+i/\tau_{5}}\right]$, where $\ell=v_F\tau$ is the mean free path. The first inequality corresponds to the condition of the normal skin effect, which states that the mean free path should be small compared to the skin depth~\cite{Landau:t8,Abrikosov:book-1988}. The second inequality defines the nonlocal regime. As we show in the main text, condition $\ell \ll \delta(\omega)$ can be rewritten as $\xi \ll1/\sqrt{\omega\tau}$. Therefore, our approximations are valid for $\xi \ll1/\sqrt{\omega\tau}$ and $\xi \gg\mbox{Re}\left[(1-i)\sqrt{1+i/(\omega\tau_{5})}\right]$. These inequalities restrict the allowed parameter region from above and below, respectively. As one can see, a longer internode scattering time is beneficial for achieving a larger parameter range where the nonlocal regime for the conditions of the normal skin effect can be realized. Notice, however, that the reduction of the parameter range at finite $\tau_5/\tau$ is weak for realistic internode scattering times $\tau_5/\tau \gtrsim 100$.

\begin{figure}[t]
\centering
\subfigure[]{\includegraphics[width=0.45\textwidth]{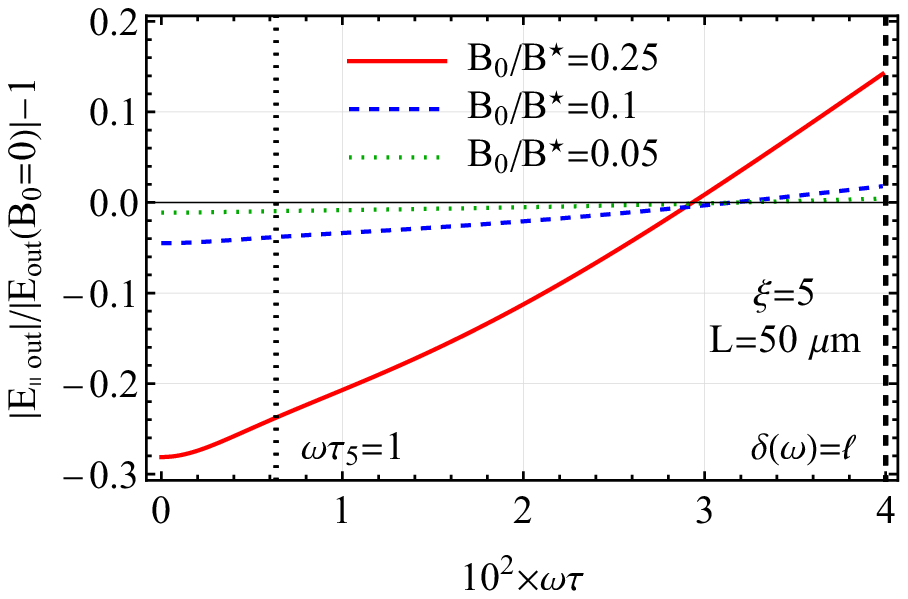}}
\hspace{0.05\textwidth}
\subfigure[]{\includegraphics[width=0.45\textwidth]{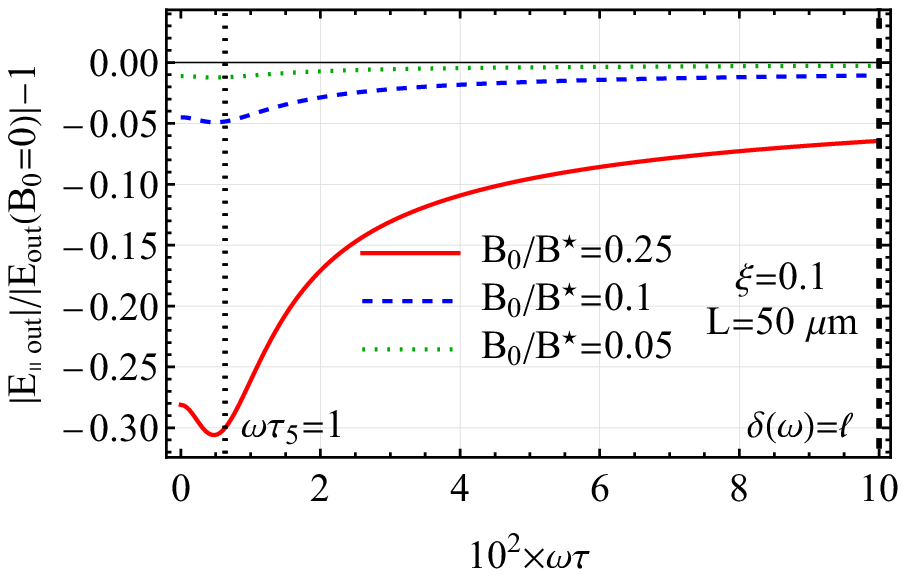}}
\vspace{-0.25cm}
\caption{
The dependence of the relative field amplitude $\left|E_{\rm \parallel out}\right|/\left|E_{\rm out}(B_0=0)\right|-1$ on frequency for a few values of the magnetic field. Panels (a) and (b) correspond to $\xi=5$ and $\xi=0.1$, respectively. Black vertical dashed lines denote the frequencies corresponding to the normal skin effect, {\sl i.e.}, $\delta(\omega)=\ell$. Black vertical dotted lines correspond to $\omega\tau_5=1$. Further, $B^\star= B_{\rm uq}\sqrt{N_{W}\tau/\tau_5}$, $B_{\rm uq}=c\mu^2/(2e \hbar v_F^2)$, $L=50~\mu\mbox{m}$, we fixed Neumann boundary conditions, and used other parameters given in the text. Notice that because of different values of $\xi$, the intranode and internode relaxation rates $1/\tau$ and $1/\tau_5$ are different in panels (a) and (b).
}
\label{fig:app-anisotropy}
\end{figure}

Finally, we present the dependence of the relative field amplitude $\left|E_{\rm \parallel out}\right|/\left|E_{\rm out}(B_0=0)\right|-1$ on frequency for a few values of the magnetic field in Figs.~\ref{fig:app-anisotropy}(a) and \ref{fig:app-anisotropy}(b) for $\xi=5$ and $\xi=0.1$, respectively. The corresponding results are obtained by using the direct approach outlined at the beginning of Sec.~S~II.C, {\sl i.e.}, we made no assumptions regarding the values of $Lq_0(\omega)$ and $Lq(\omega)$. In the calculations, we used some of the parameters of Weyl semimetal TaAs~\cite{Arnold-Felser:2016b,Zhang-Hasan-TaAs:2016}, {\sl i.e.}, $N_{W}=24$, the Fermi velocity $v_{F}\approx 3\times 10^7~\mbox{cm/s}$, the Fermi level (measured from a node) $\mu\approx 20~\mbox{meV}$, and the ratio $\tau_5/\tau\approx 158$. The intranode relaxation rate at fixed $\xi$ follows from Eq.~(\ref{app-eq:xialpha}).
In addition to the features discussed in the main text, {\sl i.e.}, a different scaling with frequency and a possibility to have an enhancement of the transmitted field, we note a nonmonotonic behavior of the relative field amplitude for $\xi\lesssim1$ with a local extremum at $\omega \tau_5=1$, see Fig.~\ref{fig:app-anisotropy}(b).

\bibliography{library-short}